\documentstyle[prd,aps,epsfig]{revtex}

\begin{document}
%
%
\def\beqa{\begin{eqnarray}}
\def\eeqa{\end{eqnarray}}
\def\be{\begin{equation}}
\def\ee{\end{equation}}
\def \wg#1{\mbox{\boldmath ${#1}$}}
\def \w#1{{\bf #1}}
\draft
\title{Quasiequilibrium sequences of synchronized and
       irrotational binary neutron stars
       in general relativity. II. Newtonian limits}
\author{Keisuke Taniguchi, Eric Gourgoulhon, and Silvano Bonazzola}
\address{D\'epartement d'Astrophysique Relativiste et de Cosmologie,
  UMR 8629 du C.N.R.S., Observatoire de Paris, \\
  F-92195 Meudon Cedex, France 
}
\date{28 May 2001}
\maketitle

\begin{abstract}

We study equilibrium sequences of close binary systems composed
of identical polytropic stars in Newtonian gravity.
The solving method is a multi-domain spectral method
which we have recently developed.
An improvement is introduced here for accurate computations
of  binary systems with stiff equation of state ($\gamma > 2$). 
The computations are performed for both cases of synchronized
and irrotational binary systems
with adiabatic indices $\gamma=3,~2.5,~2.25,~2$ and $1.8$.
It is found that the turning points of total energy
along a constant-mass sequence appear only for $\gamma \ge 1.8$ 
for synchronized binary systems
and $\gamma \ge 2.3$ for irrotational ones.
In the synchronized case, the equilibrium sequences terminate
by the contact between the two stars.
On the other hand, for irrotational binaries,
it is found that the sequences terminate at a mass shedding limit
which corresponds to a detached configuration.

\end{abstract}



\section{Introduction} \label{s:intro}

The research for equilibrium figures of binary systems
has a long history.
The pioneer work was done by Roche who treated a synchronously rotating,
incompressible ellipsoid around a gravitating point source.
Following this work, Darwin constructed a synchronized binary system
composed of double incompressible ellipsoids (see \cite{Chandra69}).
For the non-synchronized case, Aizenman calculated an equilibrium figure
of incompressible ellipsoid with internal motion orbiting
a point source companion\cite{Aizen68}.
In 1980's, the improvement of computer performance made possible
to construct equilibrium sequences of synchronized binary systems
with compressible equation of state in Newtonian gravity\cite{HachiE84}.

Since the beginning of 1990's,
the studies on binary systems gain some importance
in relation to astrophysical sources of gravitational waves.
For example, coalescing binary neutron stars are expected to be
one of the most promising sources of gravitational radiation
that could be detected by the interferometric detectors currently
in operation (TAMA300) or under construction (GEO600, LIGO, and VIRGO).
Until now, numerous theoretical studies have been done
in this research field.
Among them there are
(i)
post-Newtonian analytical treatments (e.g. \cite{BlancI98}),
(ii)
black hole perturbation\cite{TagoSTS96,TanakaTS96},
(iii)
Newtonian\cite{LaiRS93,LaiRS94,TanigN00a,TanigN00b}
and post-Newtonian\cite{LomRS97,TanigS97,ShibaT97,Tanig99}
(semi-)analytical treatments including hydrodynamical effects of stars,
(iv)
post-Newtonian hydrodynamical computations\cite{FaberR00,FaberRM01},
(v)
fully relativistic hydrodynamical treatments,
pioneered by the works of Oohara \& Nakamura (see e.g. \cite{OoharN97}),
and recently developed by
Shibata\cite{Shiba99a,Shiba99b,Shiba99c,ShibaU00},
Oohara \& Nakamura\cite{OoharN99},
and the Neutron Star Grand Challenge group\cite{Suen99,FontMST00}.
(vi)
In parallel of the dynamical studies in (v),
a quasiequilibrium formulation of the problem has been developed
\cite{BonazGM97b,Asada98,Shiba98,Teuko98}
and successfully implemented
\cite{BonazGM99a,MarroMW99,UryuE00,UryuSE00,GourGTMB01}.
The basic assumption underlying the quasiequilibirum calculations is that
the timescale of the orbit shrinking is larger than
that of the orbital revolution in the pre-coalescing state.
Consequently the evolution of the binary system can be approximated by
a succession of exactly circular orbits,
hence the name {\em quasiequilibrium}.

The first quasiequilibrium configurations of binary neutron stars
in general relativity have been obtained four years ago
by Baumgarte et al.\cite{BaumgCSST97,BaumgCSST98b},
followed by Marronetti et al.\cite{MarroMW98}.
However these computations considered synchronized binaries.
As far as coalescing binary neutron stars are concerned,
this rotation state does not correspond to physical situations,
since it has been shown that the gravitational-radiation driven evolution
is too rapid for the viscous forces to synchronize
the spin of each neutron star with the orbit\cite{Kocha92,BildsC92}
as they do for ordinary stellar binaries.
Rather, the viscosity is negligible and the fluid velocity circulation
(with respect to some inertial frame) is conserved in these systems.
Provided that the initial spins are not in the millisecond regime,
this means that close binary configurations are well approximated
by zero vorticity (i.e. {\em irrotational}) states.
Irrotational configurations are more complicated to obtain
because the fluid velocity does not vanish in the co-orbiting frame
(as it does for synchronized binaries).

We have successfully developed a numerical method to tackle this problem
and presented the first quasiequilibrium configurations
of irrotational binary neutron stars elsewhere\cite{BonazGM99a,GourGTMB01}.
The numerical technique relies on a multi-domain spectral method
\cite{BonazGM98a} within spherical coordinates.
Since then, two other groups have obtained
relativistic irrotational configurations:
(i)
Marronetti, Mathews \& Wilson\cite{MarroMW99,MarroMW00}
by means of single-domain finite difference
method within Cartesian coordinates and
(ii)
Uryu \& Eriguchi\cite{UryuE98,UryuE00,UryuSE00}
by means of multi-domain finite difference
method within spherical coordinates.

Recently, we have presented our method
in detail with numerous tests \cite{GourGTMB01} (hereafter Paper I).
Using this method, we have calculated (quasi)equilibrium sequences
of binary systems with synchronized  and irrotational rotation states.
In the present article, we will show the results in Newtonian gravity.
The results of relativistic calculations will be
given in a forthcoming article \cite{TanigGB01}.

The plan of the article is as follows.
We start by presenting the equations governing binary systems
in Newtonian gravity in Sec.~\ref{s:formul}.
Section~\ref{s:improv} is devoted to the brief explanation
about the improvement on the cases of stiff equation of state
such as $\gamma > 2$.
In Sec.~\ref{s:tests} some tests for numerical code are performed.
Then, we will show the results of evolutionary sequences of both
synchronized and irrotational binary systems constructed
by double identical stars with polytropic equation of state
in Sec.~\ref{s:results}.
Section~\ref{s:summary} contains the summary.

Throughout the present article, we use units of $G=c=1$ where
$G$ and $c$ denote the gravitational constant and speed of light,
respectively.

\section{Formulation} \label{s:formul}

Since the method which we use in the present article
for getting equilibrium configurations has already been explained
in Paper I\cite{GourGTMB01},
we will only briefly mention the basic equations and the solving procedure
in this section.

\subsection{Basic equations}

The basic equations governing the problem are as follows
(see also Sec. II.F of Paper I):
\begin{enumerate}

  \item Equation of state: \\
	We adopt a simple equation, i.e. a polytropic equation of state;

    \be
	p(H) = \kappa \, n(H)^\gamma, \label{e:eos_poly_p}
    \ee
	$p$ being the fluid pressure, $n$ the fluid baryon number density,
	$\kappa$ a constant, $\gamma$ the adiabatic index,
	and $H$ the specific enthalpy. The relation between $H$ and
	$n$ is given by
    \be
	n(H) = \left[ {\gamma-1 \over \gamma} {m_{\rm B} \over \kappa} H
	\right] ^{1/(\gamma-1)}, \label{e:eos_poly_n}
    \ee
	where $m_{\rm B}$ denotes the mean baryon mass
	($m_{\rm B}=1.66 \times 10^{-27}$ kg).

  \item Euler equation:

    \be \label{e:Euler}
	{\partial \w{v} \over \partial t} + \w{v} \cdot \vec{\nabla} \w{v}
	= -{1 \over m_{\rm B} \, n} \vec{\nabla} p - \vec{\nabla} \nu,
    \ee
	$\nu$ being the gravitational potential, and
	$\w{v}$ the velocity field in the inertial frame
	which is expressed as
    \be
	\w{v} = \w{u} + \wg{\Omega} \times \w{r} \label{e:velo_iner}
    \ee
	in terms of the velocity field in the corotating frame $\w{u}$
	and the orbital motion $\wg{\Omega} \times \w{r}$.
	In the rigidly rotating case, since there is no motion
	in the corotating frame ($\w{u}=0$), we obtain
    \be
        \w{v} = \wg{\Omega} \times \w{r}.
    \ee
	On the other hand, in the irrotational case, the
	velocity field in the inertial frame is potential, i.e.
    \be
	\w{v} = \vec{\nabla} \Psi,
    \ee
	where $\Psi$ is a scalar function.

	Under the stationary condition the Euler equation (\ref{e:Euler})
	can be integrated as
    \be
        H + \nu - {1 \over 2} (\wg{\Omega} \times \w{r})^2 = {\rm const}
	\label{e:integ_euler_co}
    \ee
        in the synchronized case and
    \be
	H + \nu + {1\over 2} (\vec{\nabla} \Psi)^2
	- (\wg{\Omega} \times \w{r})\cdot \vec{\nabla} \Psi
	= {\rm const} \label{e:integ_euler_ir}
    \ee
	in the irrotational case.

  \item Equation of continuity:

    \be
	{\partial n \over \partial t}
	+ \vec{\nabla} \cdot (n \, \w{v}) = 0.
	\label{e:continuity_general}
    \ee
	This equation is trivially satisfied by stationary rigid rotation.
	In the irrotational case, we can rewrite
	Eq. (\ref{e:continuity_general}) as
    \be \label{e:eq_psi}
	n \Delta \Psi + \vec{\nabla} n \cdot \vec{\nabla} \Psi
	= (\wg{\Omega}\times\w{r}) \cdot \vec{\nabla} n. \label{e:continuity}
    \ee

  \item Poisson equation for the gravitational field:

    \be
	\Delta \nu = 4 \pi m_{\rm B} n. \label{e:poisson}
    \ee

\end{enumerate}

\subsection{Solving procedure}

\begin{itemize}

  \item[(a)]
	First of all, we prepare the equilibrium figures of
	{\it two} spherical stars. In the present computation,
	we treat binary systems composed of equal mass stars
	with the same equation of state.
	However, we symmetrize only with respect to the orbital plane.
	Although this treatment elongates the compute time,
	we do not symmetrize the stars because we are planning to study
	binary systems composed of different mass stars
	in the series of this research.

  \item[(b)]
	The separation between the centers of the two stars is
	held fixed.
	Here we define the center as the point of the maximum enthalpy
	(or equivalently maximum density - see Eq. (\ref{e:eos_poly_n})).

  \item[(c)]
	By setting the central value of the gradient of enthalpy to be zero,
	we calculate the orbital angular velocity $\Omega$
	(see Sec. IV.D.2 of Paper I for details).

  \item[(d)]
	For irrotational configurations, the velocity potential $\Psi$
	is obtained by solving Eq. (\ref{e:continuity}).

  \item[(e)]
	Taking into account the gravitational potential from
	the companion star and the centrifugal force,
	we calculate the new enthalpy field from
	Eq.~(\ref{e:integ_euler_co}) or (\ref{e:integ_euler_ir}).

  \item[(f)]
	Using the new enthalpy, we search for the location of $H=0$;
	This defines the stellar surface
	since $H=0$ is equivalent to $p=0$
	(see Eqs. (\ref{e:eos_poly_p}) and (\ref{e:eos_poly_n})).
	Having determined the new stellar surface, we change the
	position of the inner domain boundary to make it fit with
	the stellar surface.

  \item[(g)]
	By substituting the new enthalpy in the new domain into
	Eq. (\ref{e:eos_poly_n}), the new baryon density is obtained.

  \item[(h)]
	Inserting the new baryon density into the source term of the
	Poisson equation (\ref{e:poisson}),
	we can get the new gravitational potential.

  \item[(i)]
	Finally, we compare the new enthalpy field with the old one.
	If the difference is smaller than some fixed threshold,
	the calculation is considered as having converged.
	If not, return to step (c), and continue.

\end{itemize}

During the calculation, we make the baryon mass to converge to a
fixed value by varying the central enthalpy.
Moreover, some relaxation is performed on the gravitational potential,
the centrifugal force, and the orbital angular velocity,
when such a treatment helps the iteration to converge.
The explicit procedures are detailed in Paper I.

\section{Improvement on the cases of stiff equation of state} \label{s:improv}

\subsection{Regularization technique} \label{s:regu}

The numerical method which we use in this series of research
\cite{BonazGM98a,BonazGM99a,GourGTMB01} is
a multi-domain spectral method.
In general, spectral methods lose much of their accuracy
when non-smooth functions are treated because of the so-called
Gibbs phenomenon.
This phenomenon is well known from the most familiar spectral method,
namely, the theory of Fourier series:
the Fourier coefficients ($c_n$) of a function $f$ which is of class
${\cal C}^p$ but not ${\cal C}^{p+1}$ decrease as only $1/n^p$
with increasing $n$.
In particular, if the function has some discontinuity,
its approximation by a Fourier series does not converge towards $f$
at the discontinuity point.
The {\it multi-domain} spectral method circumvents
the Gibbs phenomenon\cite{BonazGM98a}.
The basic idea is to divide the space into domains chosen so that
the physical discontinuities are located onto the boundaries between
the domains.

However, even if the circumvention of the Gibbs phenomenon has been done
by introducing  multi-domains,
there remains the Gibbs phenomenon at the boundaries between the domains
when some physical field has infinite derivative at the boundary.
For example, when we consider a star,
its surface coincides with the boundary of a domain.
If the star has a stiff equation of state, i.e. $\gamma > 2$
in terms of the adiabatic index,
the density decreases so rapidly that
$\partial n/\partial r$ diverges at the surface.
In order to recover high accuracy, we have developed a method
to regularize the density profile by extracting its diverging part
(see Sec.~IV of Ref.~\cite{BonazGM98a}).

Here we briefly summarize this method.
For a polytrope with adiabatic index $\gamma$,
the matter density $n$ behaves as $H^{1/(\gamma-1)}$
(see Eq. (\ref{e:eos_poly_n})).
In the steady state configuration, $H$ is Taylor expandable
at the neighborhood of the stellar surface because the gravitational
potential $\nu$ is Taylor expandable there
(cf. Eq. (\ref{e:integ_euler_co}) or (\ref{e:integ_euler_ir})).
Therefore $H$ vanishes as $r-R(\theta, \varphi)$,
where $R(\theta, \varphi)$ is the equation of the stellar surface.
Consequently $n$ behaves as
$n \sim [r-R(\theta, \varphi)]^{1/(\gamma-1)}$.

We introduce a known potential $\Phi_{\rm div}$ such that
$n_{\rm div}:=\Delta \Phi_{\rm div}$ has the same pathological behavior
as $n$ and such that $n-n_{\rm div}$ is a regular function
(at least more regular than $n$). We then numerically solve
\be
  \Delta \Phi_{\rm regu} = n - n_{\rm div},
\ee
where $\Phi_{\rm regu}:=\Phi-\Phi_{\rm div}$.
Consider, for instance,
\be \label{e:div_pot}
  \Phi_{\rm div} = F(\xi, \theta, \varphi)~(1-\xi^2)^{(\alpha+2)},
\ee
where $\alpha=1/(\gamma-1)$.
$F$ is an arbitrary regular function and $\xi$ is a new radial variable
such that $\xi=1$ at the surface of the star\cite{BonazGM98a}.
It is easy to see that $\Delta \Phi_{\rm div}$ has a term vanishing
at the surface with the same pathological behavior as $n$,
i.e., $(1-\xi^2)^{\alpha}$. We have indeed
\beqa
  \tilde{\Delta} \Phi_{\rm div} &=&
	\tilde{\Delta} F (1-\xi^2)^{(\alpha+2)} \nonumber \\
	&&-4(\alpha+2) \xi (1-\xi^2)^{(\alpha+1)}
		\partial_{\xi} F \nonumber \\
	&&+(\alpha+2) [ -6(1-\xi^2)^{(\alpha+1)} +4(\alpha+1) \xi^2
	(1-\xi^2)^{\alpha} ] F,
\eeqa
where $\tilde{\Delta}$ is the Laplacian computed with respect to
$(\xi, \theta, \varphi)$.

For calculational simplicity, we choose an harmonic function
for $F(\xi, \theta, \varphi)$. Then, we can write
$\Phi_{\rm div} = \sum_{l, m} a_{lm} \Phi_{lm}$, where
\be
  \Phi_{lm}:= \xi^l (1-\xi^2)^{(\alpha+2)} Y_l^m (\theta, \varphi),
\ee
$a_{lm}$ being some numerical coefficients to be determined and
$Y_l^m$ the standard spherical harmonics. We then obtain
\be
  n_{\rm div} = \sum_{l, m} a_{lm} C_l (\xi) Y_l^m (\theta, \varphi),
\ee
with
\beqa
  C_l (\xi) &=& (\alpha+2) [ -(4l+6) (1-\xi^2)^{(\alpha+1)} \xi^l
	\nonumber \\
	&&+4 (\alpha+1) \xi^{(l+2)} (1-\xi^2)^{\alpha} ].
\eeqa
We now have to determine the values of the coefficients $a_{lm}$
which give the most regular function $n_{\rm regu}:=n-n_{\rm div}$.
The criterion which seems to give the best results is the following one.
We expand both $n$ and $n_{\rm div}$
as truncated series of spherical harmonics
$Y_l^m (\theta, \varphi)$ and Chebyshev polynomial $T_i (\xi)$,
\beqa
  &&n (\xi, \theta, \varphi) =\sum_{i,l,m=0}^{I,L,M} n_{ilm} T_i (\xi)
	Y_l^m (\theta, \varphi), \\
  &&n_{\rm div} (\xi, \theta, \varphi) = \sum_{i,l,m=0}^{I,L,M}
	a_{lm} C_{li} T_i (\xi) Y_l^m (\theta, \varphi),
\eeqa
where we expand each of the function $C_l (\xi)$ in a Chebyshev series as
\be
  C_l (\xi) =\sum_{i=0}^I C_{li} T_i (\xi).
\ee
The values of $a_{lm}$ is computed in such a way that the $I$th coefficient
of the truncated series of $n_{\rm regu}$ vanishes:
\be
  a_{lm} = {n_{Ilm} \over C_{lI}}.
\ee

By means of the above procedure, we eliminate in $n$ the pathological
term which vanishes as $(1-\xi^2)^{\alpha}$ but we introduce another
pathological term proportional to $(1-\xi^2)^{\alpha+1}$.
However, the divergence occurs in a higher order derivative of this term
so that it has a much weaker effect on the accuracy of the results.
The method can be improved by taking
\beqa
  \Phi_{\rm div} &=& F (\xi, \theta, \varphi) (1-\xi^2)^{(\alpha+2)}
	[a_1 +a_2 (1-\xi^2) \nonumber \\
	&&+a_3 (1-\xi^2)^2 +\cdots +a_K (1-\xi^2)^{K-1} ]
\eeqa
instead of Eq. (\ref{e:div_pot}).
The coefficients $a_K$ are chosen in such a way that the first, second,
..., $K$th derivatives of $n_{\rm regu}$ vanish at $\xi=1$.
Let us call $K$ the regularization degree of the procedure.

The explicit procedure to determine the coefficients $a_K$ is as follows:
Let us consider the case of $K=2$ and choose an harmonic function
for $F(\xi, \theta, \varphi)$ for example.
In this case, we need two diverging terms, $n_{{\rm div}:k}~(1\le k \le K)$,
where we define
\be
  n_{{\rm div}:k} :=\sum_{l, m} a_{lm}^k C_l^k (\xi) Y_l^m (\theta, \varphi),
\ee
with
\beqa
  C_l^k (\xi) &=& (\alpha+k+1) [ -(4l+6) (1-\xi^2)^{(\alpha+k)} \xi^l
	\nonumber \\
	&&+4 (\alpha+k) \xi^{(l+2)} (1-\xi^2)^{\alpha+k-1} ].
\eeqa
First of all, we expand $n$, $n_{{\rm div}:1}$ and $n_{{\rm div}:2}$ by
$Y_l^m (\theta, \varphi)$ and $T_i (\xi)$.
Then, we can express $C_l^k (\xi)$ as an expansion in a Chebyshev series,
\be
  C_l^k (\xi) =\sum_{i=0}^I C_{li}^k T_i (\xi).
\ee
Finally, we solve the simultaneous equations
\beqa
  n_{Ilm} &=& a_{lm}^1 C_{lI}^1 + a_{lm}^2 C_{lI}^2, \\
  n_{I-1lm} &=& a_{lm}^1 C_{lI-1}^1 + a_{lm}^2 C_{lI-1}^2,
\eeqa
so that the $I$th and $(I-1)$th coefficients of the truncated
series of $n_{\rm regu}$ vanish.

\subsection{Illustration and test}

\begin{figure}
\vspace{0.3cm}
  \centerline{ \epsfig{figure=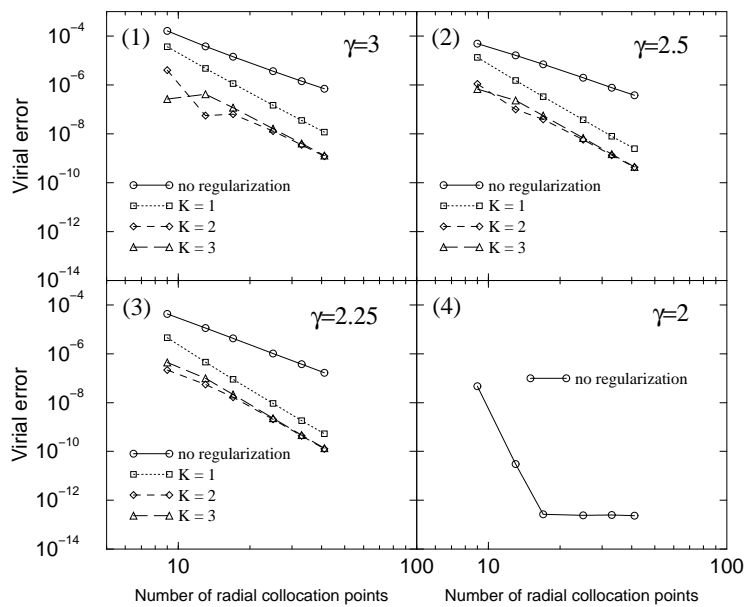,height=8cm} }
\vspace{0.3cm}
\caption[]{\label{fig:virial_regu}
Relative error in the virial theorem for a spherical star
as a function of the number of radial collocation points $N_r$.
Solid line with open circle, dotted with open square,
dashed with open diamond, and long-dashed with open triangle denote
the results without regularization, with regularization $K=1$,
$2$, and $3$, respectively.
}
\end{figure}

To validate the above regularization technique,
we investigate the improvement
in the relative error in the virial theorem for a spherical static star,
which is defined as
\be \label{e:vir_error_static}
  {\rm Virial~error} = {|W +3P| \over |W|},
\ee
$W$ being the gravitational potential energy and
$P$ the volume integral of the fluid pressure.
We give the values of the virial error in Fig. \ref{fig:virial_regu}.
In this figure we show the cases of (1) $\gamma=3$, (2) $\gamma=2.5$,
(3) $\gamma=2.25$, and (4) $\gamma=2$, and in each small figure
except for the case of $\gamma=2$ for which no Gibbs phenomenon occurs,
we have drawn four lines: ``no regularization'',
regularization degree $K=1$, 2, and 3.
It is found from these plots that the regularization dramatically
improves the accuracy.
However, the improvement by the regularization saturates at $K=2$.
Note here that since there is no Gibbs phenomenon
in the calculation of a spherical star in the case of $\gamma=2$,
the virial error decreases rapidly (exponential decay - evanescent error)
with the increase of the number of radial collocation point $N_r$,
and reaches the minimum value
due to the finite number of digits (15=double precision)
used in the numerical computation.

Considering the saturation of the virial error,
we will choose the regularization degree $K=2$
in the following numerical computations for $\gamma>2$.
We note that since the matter density $n$ behaves as $H^{1/(\gamma-1)}$,
the derivatives of $n$ of order higher than the second one diverge
at the stellar surface even if $\gamma<2$.
Therefore we will also use the regularization ($K=2$)
for the cases of $\gamma<2$.

\section{Tests of the numerical code} \label{s:tests}

Numerous tests have been already performed in the previous paper
in the irrotational case (see Sec. V.B of Paper I). In particular
a direct comparison with the numerical results of Uryu \& Eriguchi 
\cite{UryuE98} has been performed and the result presented in
Table~I of Paper~I. Here we will focus on some tests in the synchronized case,
not presented in Paper~I. We will also show the 
relative error on the virial theorem along an evolutionary sequence
in both cases, extending the test presented in Paper~I (Fig.~7)
to adiabatic indices different from 2. 

\subsection{Comparison with analytical solutions}

\begin{figure}
\vspace{0.3cm}
  \centerline{ \epsfig{figure=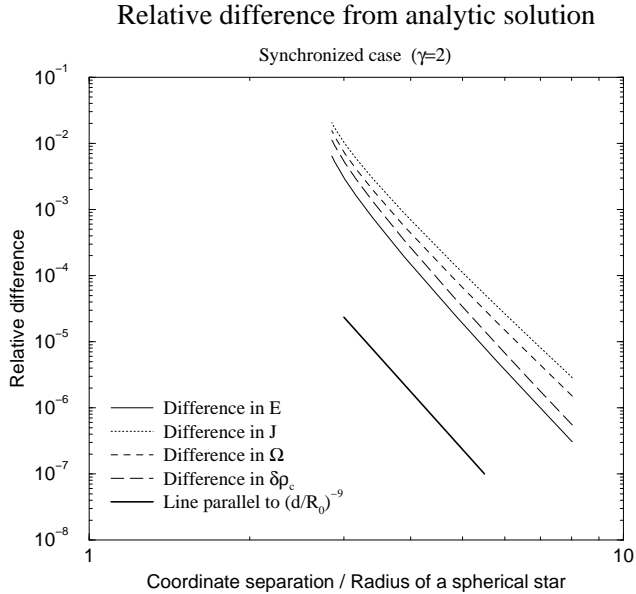,height=8cm} }
\vspace{0.3cm}
\caption[]{\label{fig:rel_diff}
Relative differences in total energy $E$, total angular momentum
$J$, orbital angular velocity $\Omega$, and relative change in central
baryon density $\delta \rho_c$ when comparing the numerical solution
with Taniguchi \& Nakamura's approximate analytic solution \cite{TanigN01}
along an equilibrium sequence in the synchronized case with $\gamma=2$.
The horizontal axis denotes logarithmically
$d/R_0$, where $d$ is the separation between the two stellar centers,
and $R_0$ the stellar radius at infinite separation.
The thick solid line is a reference one in order to check the inclinations
of the results easily.}
\end{figure}

In order to investigate the discrepancy between the results from
the numerical code and those from Taniguchi \& Nakamura's analytic
solution\cite{TanigN01} in the synchronized case with $\gamma=2$,
we present the relative differences
on global quantities as functions of the separation in a log-log plot
in Fig. \ref{fig:rel_diff} (in the same way as Fig.~12 of Paper I
for irrotational configurations).
The relative differences are defined as follows:
\beqa
  &&{E_{\rm num} -E_{\rm ana} \over GM^2/R_0}, \\
  &&{J_{\rm num} -J_{\rm ana} \over Md^2 \Omega_{\rm Kep}/2}, \\
  &&{\Omega_{\rm num} -\Omega_{\rm ana} \over \Omega_{\rm Kep}}, \\
  &&|\delta \rho_{c: {\rm num}} -\delta \rho_{c: {\rm ana}}|,
\eeqa
where $\Omega_{\rm Kep}$ is the Keplerian velocity for point mass particles
\be
  \Omega_{\rm Kep}:= \Bigl( {2GM \over d^3} \Bigr)^{1/2}.
\ee
In the Taniguchi \& Nakamura's analytic solution \cite{TanigN01},
the global quantities are expanded up to $O[(R_0/d)^6]$.
After subtracting these analytic solutions from the numerical one,
only the terms of order higher than $O[(R_0/d)^6]$ should remain.
Indeed, we can see from Fig. \ref{fig:rel_diff} that
the discrepancies between numerical and analytical solutions for
the energy $E$ and the relative change in central density $\delta \rho_c$
are both higher than $O[(R_0/d)^8]$,
and those for the angular momentum $J$ and the orbital angular velocity
$\Omega$ are both higher than $O[(R_0/d)^7]$. This means that the
numerical solution and the analytical one agree up to the
accuracy of this latter ($O[(R_0/d)^6]$).

\subsection{Virial theorem}

\begin{figure}
\vspace{0.3cm}
  \centerline{ \epsfig{figure=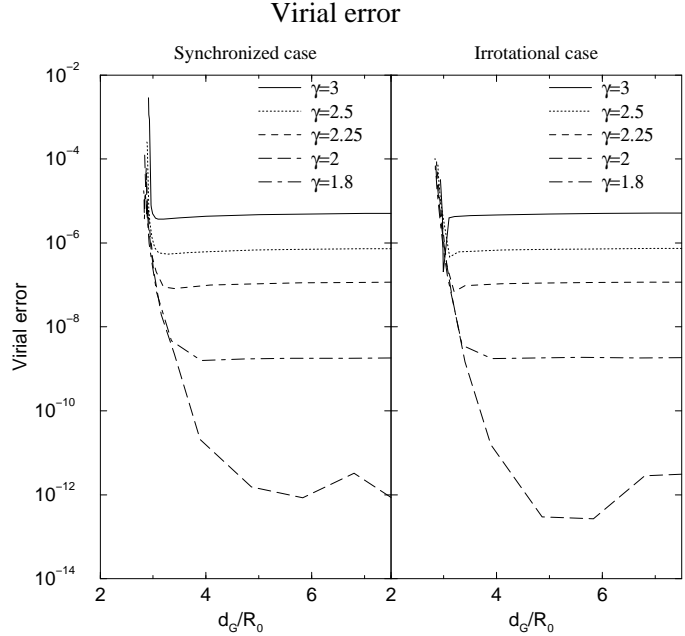,height=8cm} }
\vspace{0.3cm}
\caption[]{\label{fig:virial}
Relative error in the virial theorem along an evolutionary sequence.
The left panel is for synchronized binaries,
and the right one for irrotational binaries.
Solid, dotted, dashed, long-dashed, and dot-dashed lines denote
the cases $\gamma=3$, $2.5$, $2.25$, $2$, and $1.8$, respectively.
The horizontal axis denotes $d_G/R_0$, where $d_G$ is the separation
between the centers of masses of the two stars, and $R_0$ the stellar
radius at infinite separation.}
\end{figure}

One of the best indicators of the accuracy of numerical
solutions for binary systems in Newtonian gravity is the relative error
in the virial theorem.
This error is defined as
\be \label{e:vir_error}
  {\rm Virial~error} ={|2T +W +3P| \over |W|},
\ee
where $T$, $W$, and $P$ denote respectively the kinetic energy of
the binary system, its gravitational potential energy
and the volume integral of the fluid pressure.
By virtue of the virial theorem, the quantity defined by
Eq.~(\ref{e:vir_error}) should be zero for an exact solution.
We show it in Fig.~\ref{fig:virial}.
This Figure can be considered as an extension to that presented
in Paper~I (Fig.~7) to adiabatic indices different from 2.
It also contains the synchronized case. 
We can see from Fig.~\ref{fig:virial} that the code is quite accurate as
${\rm Virial~error}<10^{-5}$ for $d_G/R_0 \ge 3$ even if $\gamma=3$,
where $d_G$ denotes the separation between two centers of masses
of each star.

\section{Results} \label{s:results}

\begin{figure}
\vspace{0.3cm}
  \centerline{ \epsfig{figure=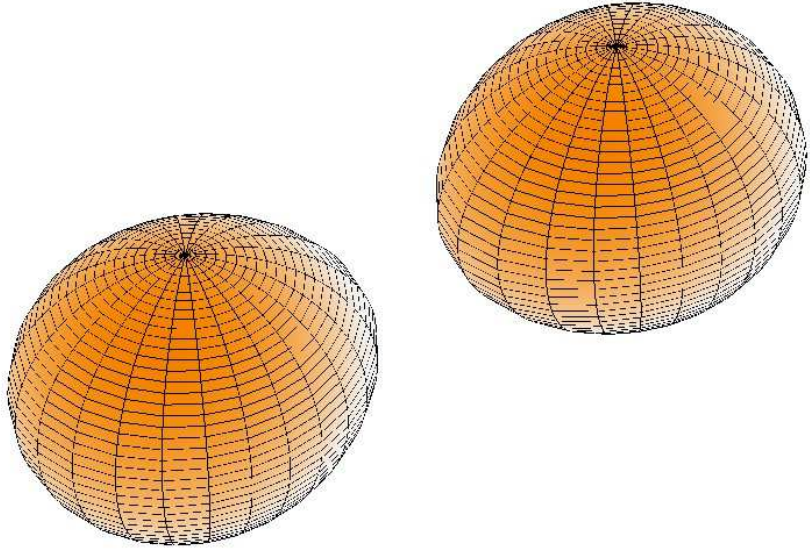,height=8cm} }
\vspace{0.3cm}
\caption[]{\label{f:3d_sync}
Synchronized binary at the point of minimum energy and angular momentum
along a constant-mass sequence (``last stable orbit''), for an adiabatic
index $\gamma=3$. The grids on the surfaces corresponds to the
collocation points in $(\theta,\varphi)$ of the spectral method.}
\end{figure}

\begin{figure}
\vspace{0.3cm}
  \centerline{ \epsfig{figure=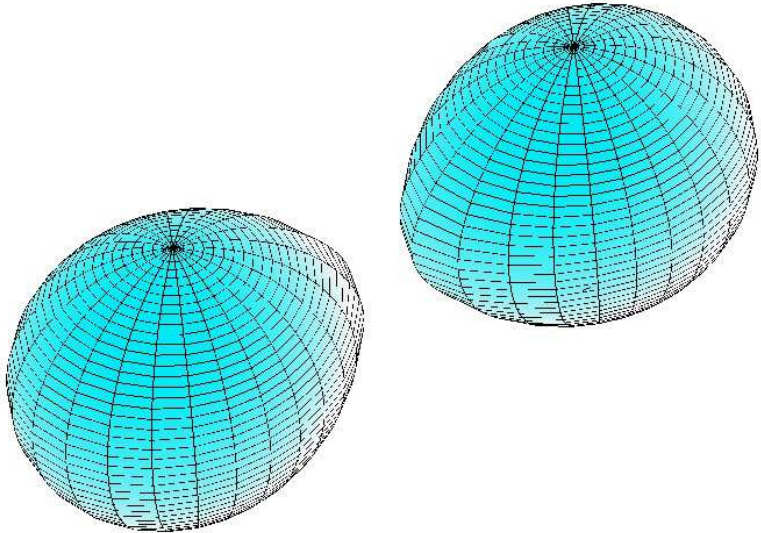,height=8cm} }
\vspace{0.3cm}
\caption[]{\label{f:3d_irrot}
Same as Fig.~\ref{f:3d_sync} but for an irrotational binary.}
\end{figure}

Using the numerical method explained in the above sections
and in Paper I\cite{GourGTMB01},
we have constructed equilibrium sequences of both synchronized and
irrotational binary systems in Newtonian gravity.
Here, we consider only the case of binary systems composed of
stars with equal mass and equation of state.
We use 3 domains (one for the fluid interior) for each star and
the following number of spectral coefficients:
$N_r \times N_{\theta} \times N_{\varphi} = 33 \times 25 \times 24$
in each domain.
A view of the configuration at the energy turning point 
(Sec.~\ref{s:turning_points} below)
along a sequence with a $\gamma=3$ EOS is shown in Fig.~\ref{f:3d_sync}
for synchronized binaries, and in Fig.~\ref{f:3d_irrot} for 
irrotational one. 
We are interested in the turning point of total energy
(and/or total angular momentum)
because it corresponds to the onset of secular instability
in the synchronized case\cite{BaumgCSST98a}
and that of dynamical instability in the irrotational one
at least for the ellipsoidal approximation\cite{LaiRS93}.

\subsection{Equilibrium sequences} \label{s:equil_seq}

Our results for equilibrium constant-mass sequences 
(evolutionary sequences) with adiabatic indices
$\gamma=3,~2.5,~2.25,~2$ and 1.8 are presented 
in Tables~\ref{table1} and \ref{table2}.
In these tables, $d$ denotes the separation between
the centers of the two stars.
Let us recall that the center of a star is defined as
the point of maximum enthalpy (or equivalently maximum density).
On the other hand, $d_G$ denotes the orbital separation between centers
of masses of two stars.
$R_0$, $a_1$, $a_2$, $a_3$, and $a_{1,{\rm opp}}$ are
the radius of a spherical star of same mass,
the radius parallel to x-axis toward the companion star,
the radius parallel to y-axis,
the radius parallel to z-axis, and
the radius parallel to x-axis opposite to the companion star.
The $(x, y, z)$ axes are the same as in Fig. 1 of Paper I.
$\rho_c$ and $\rho_{c0}$ indicate the central density of a star
and that of a spherical star of same mass.
The normalized quantities $\bar{\Omega}$, $\bar{J}$, and $\bar{E}$ are
defined by 
\beqa
  \bar{\Omega} &:=& {\Omega \over (\pi G \rho_0)^{1/2}}, \\
  \bar{J} &:=& {J \over (G M^3 R_0)^{1/2}}, \\
  \bar{E} &:=& {E \over GM^2/R_0}, \\
\eeqa
where $\Omega$, $J$, and $E$ denote respectively the orbital angular velocity,
the total angular momentum, and the total energy, 
and $\rho_0$ is the averaged density of a spherical star of the same mass:
\be
  \rho_0 :={3 M \over 4 \pi R_0^3}.
\ee

Also listed in Tables~\ref{table1} and \ref{table2} is the ratio
\be
  \chi := {(\partial H/\partial r)_{\rm eq,comp} \over
  		(\partial H/\partial r)_{\rm pole}},
	\label{e:chi}
\ee
where $(\partial H/\partial r)_{\rm eq,comp}$ 
[resp. $(\partial H/\partial r)_{\rm pole}$]
stands for the radial derivative of the enthalpy
at the point on the stellar surface located in the orbital plane and
looking toward the companion star [resp. at the intersection between
the surface and the axis perpendicular to the orbital plane and going 
through the stellar center ($z$ axis)]. 
This quantity is useful because the mass shedding limit (``Roche limit'')
corresponds to $\chi=0$ (cf. Sec.~IV.E of Paper~I). When $\chi=0$, 
an angular point (cusp) appears at the equator of the star in the
direction to the companion. 

An equilibrium sequence terminates either by a contact configuration, 
which corresponds to $d/a_1=2$ or by the mass shedding limit given
by $\chi=0$. Note that these two conditions are not exclusive, as
one can have $\chi=0$ at contact. 
By means of our numerical method, it is difficult to get exactly
such configurations, so that the final points in Tables~\ref{table1} 
and \ref{table2} are close to, but not exactly equal to the 
real end points $d/a_1=2$ or $\chi=0$.

For synchronized binaries, the real end point is the contact one
\cite{HachiE84}. In that configuration, the surface of each star
has a cusp at the contact point. However, our numerical multi-domain
method assumes that the boundary between each domain is a differentiable
surface. Therefore, for computing very close configurations, we stop the 
adaptation of the domain to the surface of the star when $\chi<0.3 \sim 0.35$
(see Sec.~IV.E of Paper~I). For synchronized binaries, the use of adapted
domain is not essential, since no boundary condition is set at the stellar
surface. The last lines for each $\gamma$ in Table \ref{table1} are obtained
in this way\footnote{Due to the overlapping of the external domains for contact
configurations, it was not possible to compute these latter with
high accuracy. This is why we stopped Table~\ref{table1} slightly 
before $d/a_1=2$}.  

On the other hand, in the irrotational case, leaving the adaptation
of the domain to the stellar surface, results in some numerical error
since the solving method for the equation governing the velocity
potential [Eq.~(\ref{e:eq_psi})] assumes that the domain boundary
coincide with the stellar surface (see Appendix~B of Paper~I). 
Therefore we stop the calculations at some points
which are very close to the cusp points but slightly separated.

The symbol $\dagger$ in Tables~\ref{table1} and \ref{table2}
indicates the points
of minimum total energy (and total angular momentum) along the sequence, 
also called {\em turning points}. 
In the synchronized case, the minimum points exist for $\gamma \ge 2$,
and in the irrotational case, they do for $\gamma \ge 2.5$.
In both synchronized and irrotational cases, the minimum points
of total energy and total angular momentum coincide with each other.
It is worth to note here that we cannot exclude the possibility
of existence of minimum points for $\gamma < 2$ in the synchronized case
and for $\gamma < 2.5$ in the irrotational one,
because we do not calculate up to contact points.
However, we suspect that the critical values of the existence of minimum
points are $\gamma \sim 1.8$ in the synchronized case
and $\gamma \sim 2.3$ in the irrotational one.
The detailed discussions are given later (see Sec.~\ref{s:turning_points}).

\begin{figure}
\vspace{0.3cm}
  \centerline{ \epsfig{figure=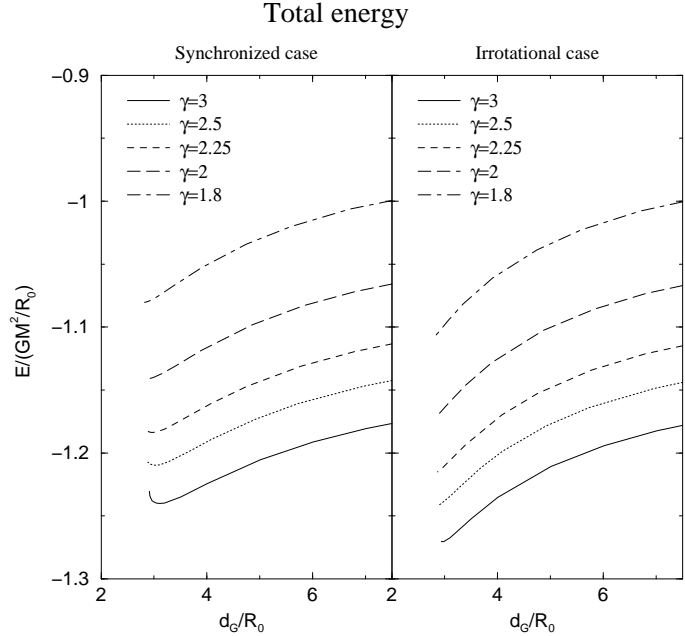,height=8cm} }
\vspace{0.3cm}
\caption[]{\label{fig:energy}
Total energy along an evolutionary sequence.
The left panel is for synchronized binaries,
and the right one for irrotational binaries. 
Solid, dotted, dashed, long-dashed, and dot-dashed lines denote
the cases of $\gamma=3$, $2.5$, $2.25$, $2$, and $1.8$, respectively.}
\end{figure}

\begin{figure}
\vspace{0.3cm}
  \centerline{ \epsfig{figure=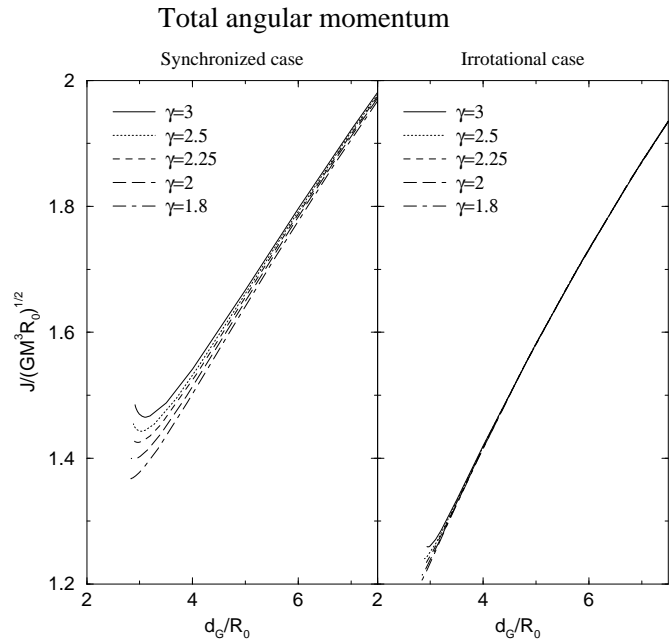,height=8cm} }
\vspace{0.3cm}
\caption[]{\label{fig:angmom}
Same as Fig. \ref{fig:energy} but for the total angular momentum.}
\end{figure}

\begin{figure}
\vspace{0.3cm}
  \centerline{ \epsfig{figure=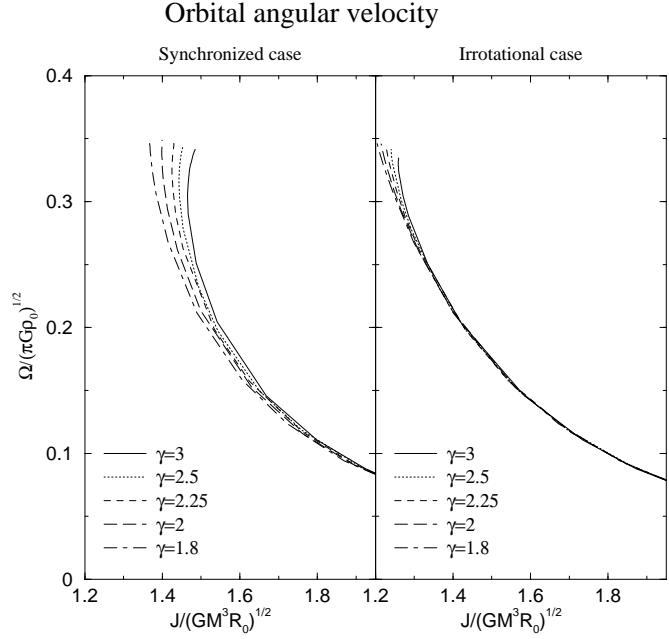,height=8cm} }
\vspace{0.3cm}
\caption[]{\label{fig:omega}
Orbital angular velocity as a function of total angular momentum
along an evolutionary sequence.
The lines have the same meaning as in Fig. \ref{fig:energy}.}
\end{figure}

\begin{figure}
\vspace{0.3cm}
  \centerline{ \epsfig{figure=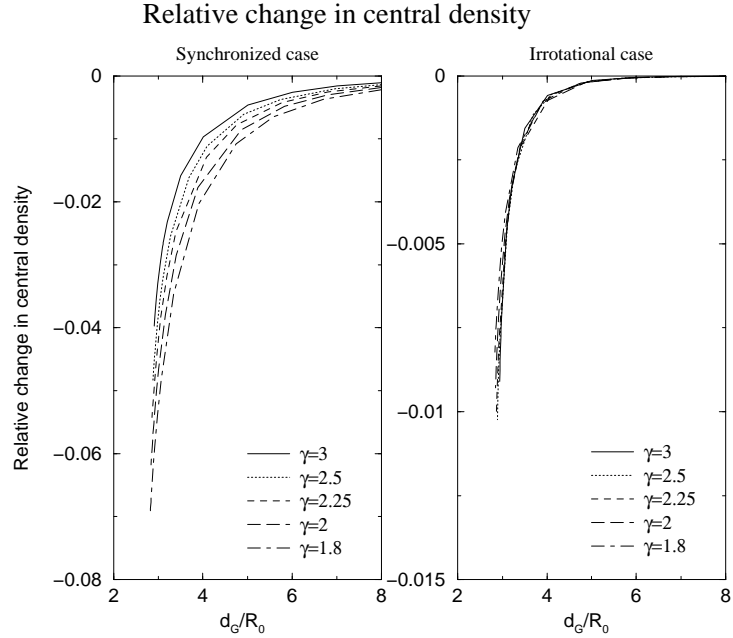,height=8cm} }
\vspace{0.3cm}
\caption[]{\label{fig:change}
Same as Fig. \ref{fig:energy} but for the relative change in central
baryon density. Note that the two vertical scales are different.}
\end{figure}

We show the total energy, the total angular momentum,
the orbital angular velocity, and the relative change in central density
along a constant mass sequence
in Figs. \ref{fig:energy} -- \ref{fig:change}.
We can see from figures of total energy and total angular momentum
that the values for synchronized systems are larger than those for
irrotational ones.
This is because the effect of spin of each star in the synchronized case
is larger than that in the irrotational one.
Such a spin produces not only the direct effects like spin energy or
spin angular momentum but also larger deformations which results in
larger quadrupole moments.
We can see clearly a turning point in energy and angular momentum curves
for large values of $\gamma$
on Figs.~\ref{fig:energy} -- \ref{fig:omega}. 
This feature will be discussed in Sec.~\ref{s:turning_points}.

It is found from Fig. \ref{fig:change} that the central density
decreases in both cases of synchronized and irrotational.
The decrease of the central density in the synchronized case
is about one order larger than that in the irrotational one.
These behaviors are analytically known.
For the synchronized case, Chandrasekhar obtained the lowest order change
in the central density about 70 years ago\cite{Chandra33}
and Taniguchi \& Nakamura have calculated the higher order 
change\cite{TanigN01}, and Taniguchi \& Nakamura have also shown it
for the irrotational case\cite{TanigN00a,TanigN00b}.

\begin{figure}
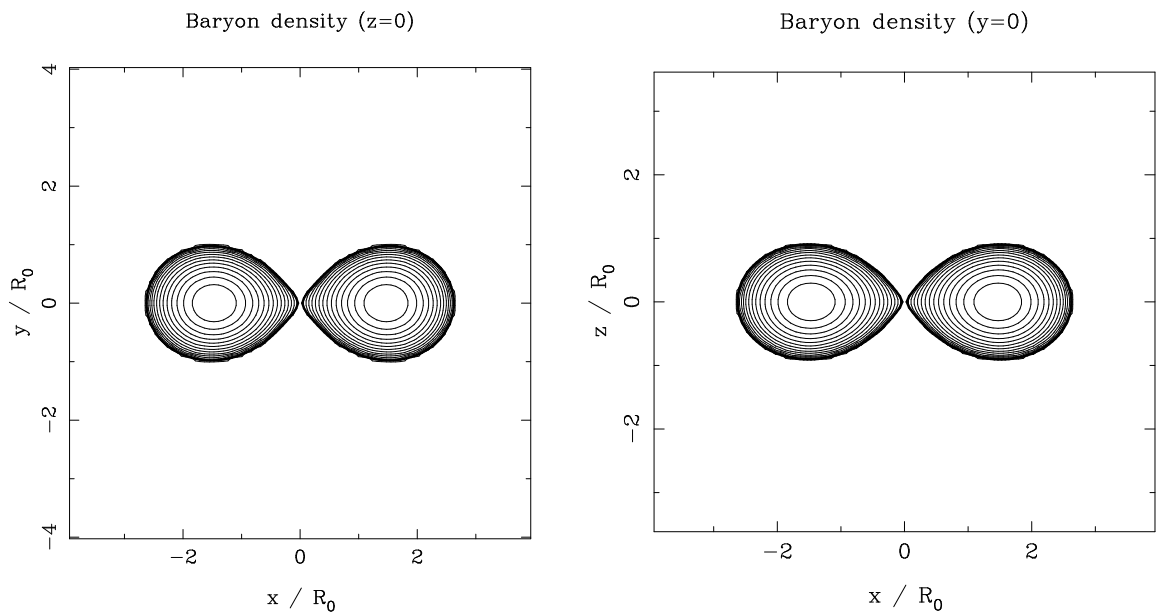

  \centerline{ \epsfig{figure=rbxy_f_cg3.eps,height=8cm}
	\hspace{0.5cm} \epsfig{figure=rbxz_f_cg3.eps,height=8cm} }
\vspace{0.3cm}
\caption[]{\label{fig:rho_bin_cg3}
Isocontour of the baryon density of synchronized binaries with $\gamma=3$
when the separation is $d/R_0=2.941$.
The plots are cross sections of $Z=0$ and $Y=0$ planes.
The thick solid lines denote the stellar surface.
The small rough on the stellar surface is an artifact of the graphical software.
}
\end{figure}

In Figs. \ref{fig:rho_bin_cg3} -- \ref{fig:rho_bin_cg18},
we show isocontours of baryon density in the synchronized case
with $\gamma=3$, $2$ and $1.8$, respectively.
These figures correspond to the configurations in the last lines
(for each adiabatic index $\gamma$) in Table \ref{table1}.
Note here that the small rough on the stellar surface of $\gamma=3$
in Fig. \ref{fig:rho_bin_cg3} is an artifact of the graphical software. 
The solved figure is a completely smooth one, thanks to the technique
discussed in Sec.~\ref{s:regu}.

\begin{figure}
  \centerline{ \epsfig{figure=rbxy_f_cg2.eps,height=8cm}
	\hspace{0.5cm} \epsfig{figure=rbxz_f_cg2.eps,height=8cm} }
\vspace{0.3cm}
\caption[]{\label{fig:rho_bin_cg2}
Same as Fig. \ref{fig:rho_bin_cg3} but for $\gamma=2$
with the separation $d/R_0=2.849$.}
\end{figure}

\begin{figure}
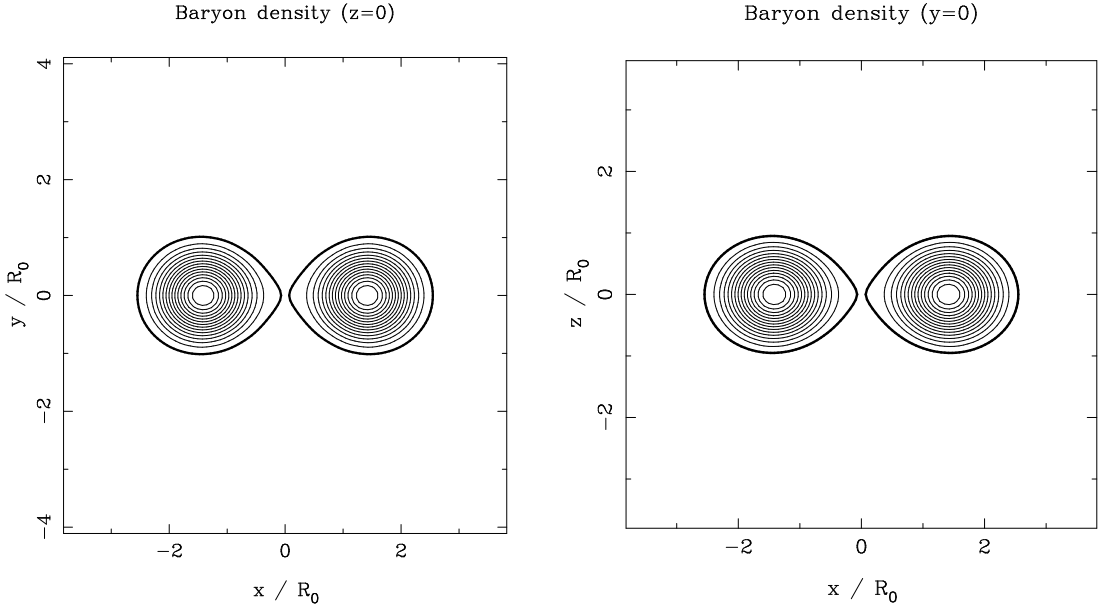

  \centerline{ \epsfig{figure=rbxy_f_cg18.eps,height=8cm}
	\hspace{0.5cm} \epsfig{figure=rbxz_f_cg18.eps,height=8cm} }
\vspace{0.3cm}
\caption[]{\label{fig:rho_bin_cg18}
Same as Fig. \ref{fig:rho_bin_cg3} but for $\gamma=1.8$
with the separation $d/R_0=2.828$.}
\end{figure}

\begin{figure}
  \centerline{ \epsfig{figure=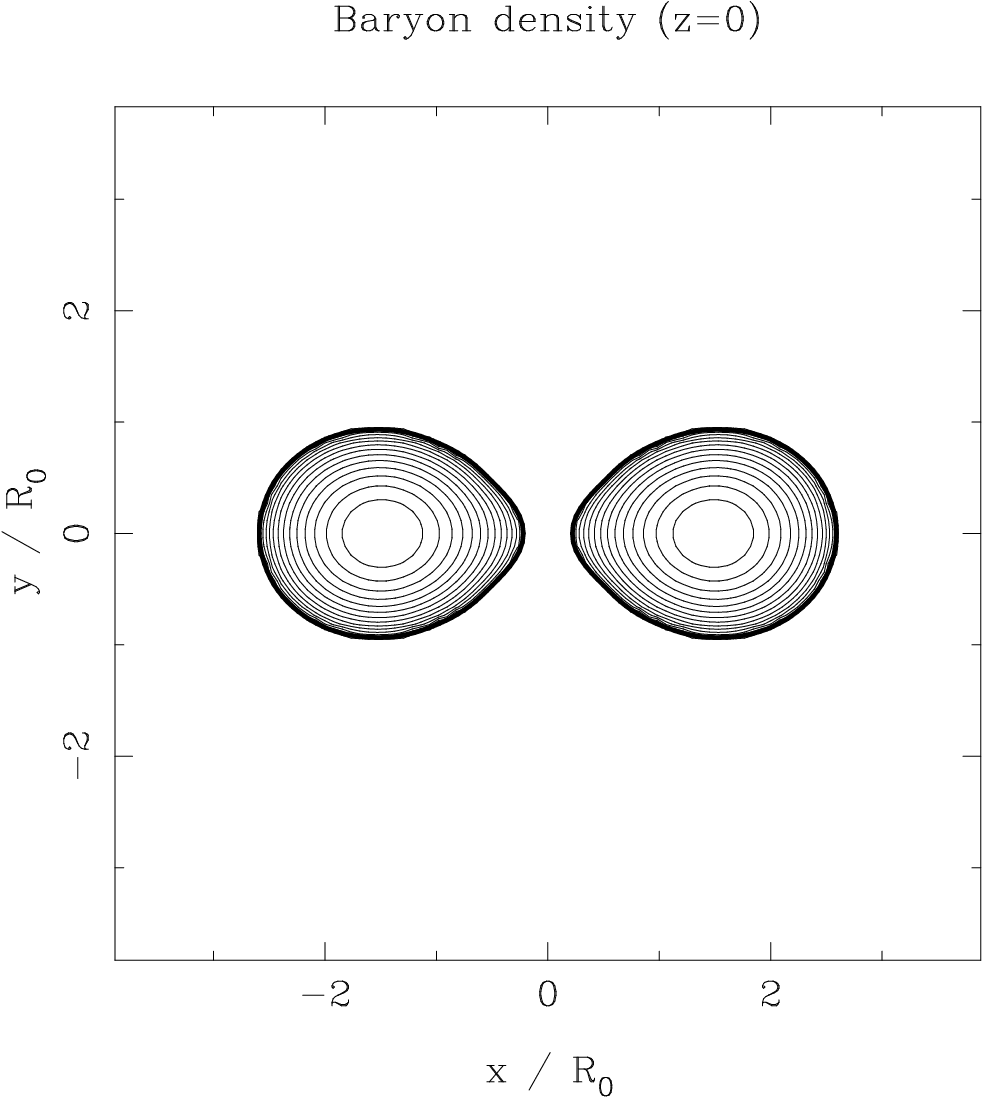,height=8cm}
	\hspace{0.5cm} \epsfig{figure=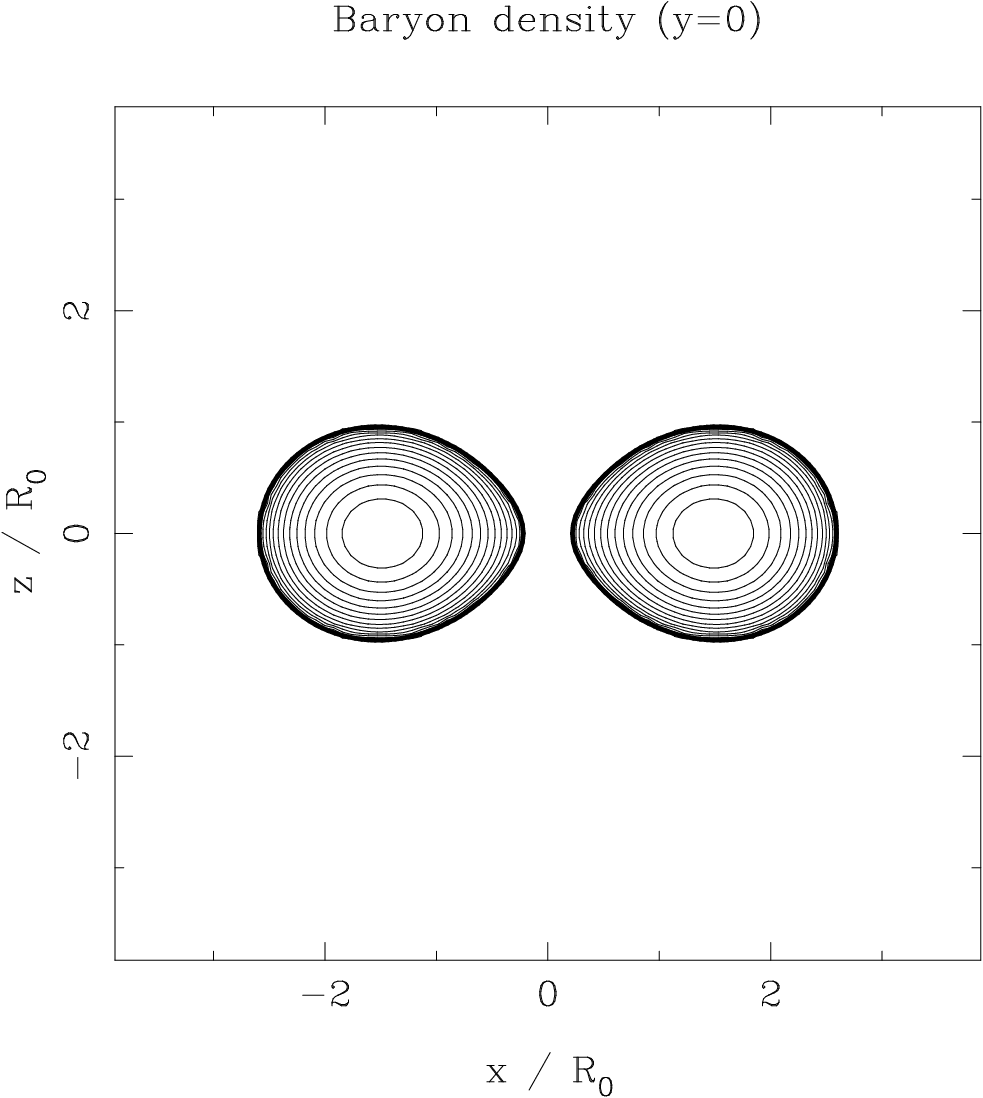,height=8cm} }
\vspace{0.3cm}
\caption[]{\label{fig:rho_bin_ig3}
Isocontour of the baryon density in the irrotational case with $\gamma=3$
when the separation is $d/R_0=2.976$.
The plots are cross sections of $Z=0$ and $Y=0$ planes.
The thick solid lines denote the stellar surface.
The small rough on the stellar surface is an artifact of the graphical
software.
}
\end{figure}

Isocontours of baryon density with $\gamma=3$, $2$ and $1.8$
for irrotational binaries are shown 
in Figs. \ref{fig:rho_bin_ig3} -- \ref{fig:rho_bin_ig18}.
These configurations correspond to the semi-final lines
(for each adiabatic index $\gamma$) in Table \ref{table2}.
Again note that the small rough on the stellar surface of $\gamma=3$
in Fig. \ref{fig:rho_bin_ig3} is an artifact of the graphical software and
that the solved figure is a completely smooth one, thanks to the technique
discussed in Sec.~\ref{s:regu}.
We do not depict the configurations of the last lines of Table~\ref{table2}
for the following reason. 
As mentioned above and in Paper I,
we make the boundary the inner domain fit with the stellar surface.
This procedure is essential to accurately solve equilibrium figures
in the irrotational case (Appendix~B of Paper~I).
However, due to the apparition of the cusp on the stellar surface, 
we can no longer adapt the domain to that surface
because all quantities are expressed by summation of a finite number of
differentiable functions. For very close configurations, just prior to 
the apparition of the cusp, the surface is highly distorted so that
there appear unphysical oscillations when using finite series of
differentiable functions (Gibbs phenomenon). 
In the last lines for each $\gamma$ in Table \ref{table2},
such oscillations start to be seen, although they do not alter
appreciably the global quantities. 
Therefore to avoid any misunderstanding, 
we choose not to plot the final lines in Table \ref{table2}, 
although we listed them.

\begin{figure}
  \centerline{ \epsfig{figure=rbxy_sf_ig2.eps,height=8cm}
	\hspace{0.5cm} \epsfig{figure=rbxz_sf_ig2.eps,height=8cm} }
\vspace{0.3cm}
\caption[]{\label{fig:rho_bin_ig2}
Same as Fig. \ref{fig:rho_bin_ig3} but for $\gamma=2$
with the separation $d/R_0=2.917$.}
\end{figure}

\begin{figure}
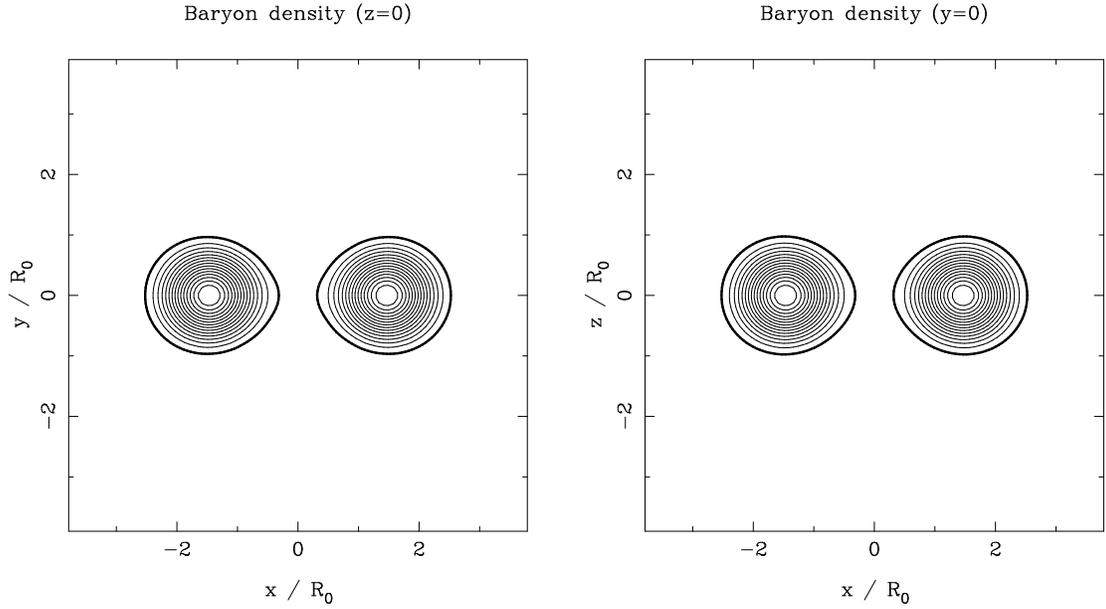

  \centerline{ \epsfig{figure=rbxy_sf_ig18.eps,height=8cm}
	\hspace{0.5cm} \epsfig{figure=rbxz_sf_ig18.eps,height=8cm} }
\vspace{0.3cm}
\caption[]{\label{fig:rho_bin_ig18}
Same as Fig. \ref{fig:rho_bin_ig3} but for $\gamma=1.8$
with the separation $d/R_0=2.932$.}
\end{figure}

\begin{figure}
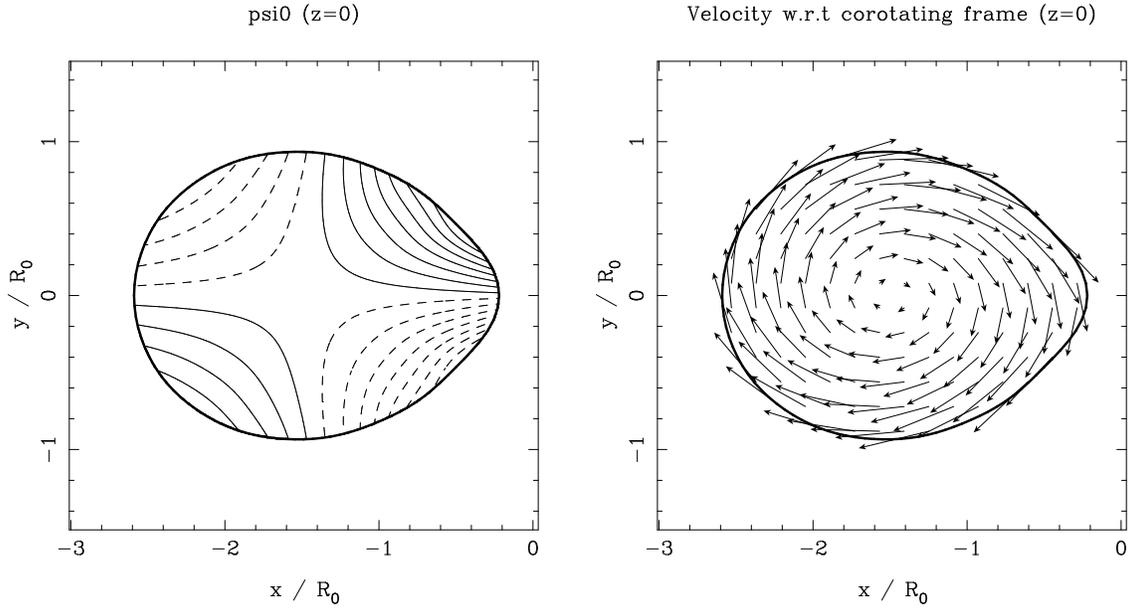

  \centerline{ \epsfig{figure=psi0_sf_ig3.eps,height=8cm}
	\hspace{0.5cm} \epsfig{figure=vs_corot_sf_ig3.eps,height=8cm} }
\vspace{0.3cm}
\caption[]{\label{fig:velo_ig3}
Contour of velocity potential $\Psi_0$ (left-hand side)
and internal velocity field $\w{u}$ (right-hand side)
with respect to the co-orbiting frame in the orbital plane
in the irrotational case with $\gamma=3$ when the separation $d/R_0=2.976$.
The thick solid lines denote the stellar surface.
The thin solid and dashed lines in the figure of velocity potential
(left-hand side) denote positive and negative values, respectively.
}
\end{figure}

\begin{figure}
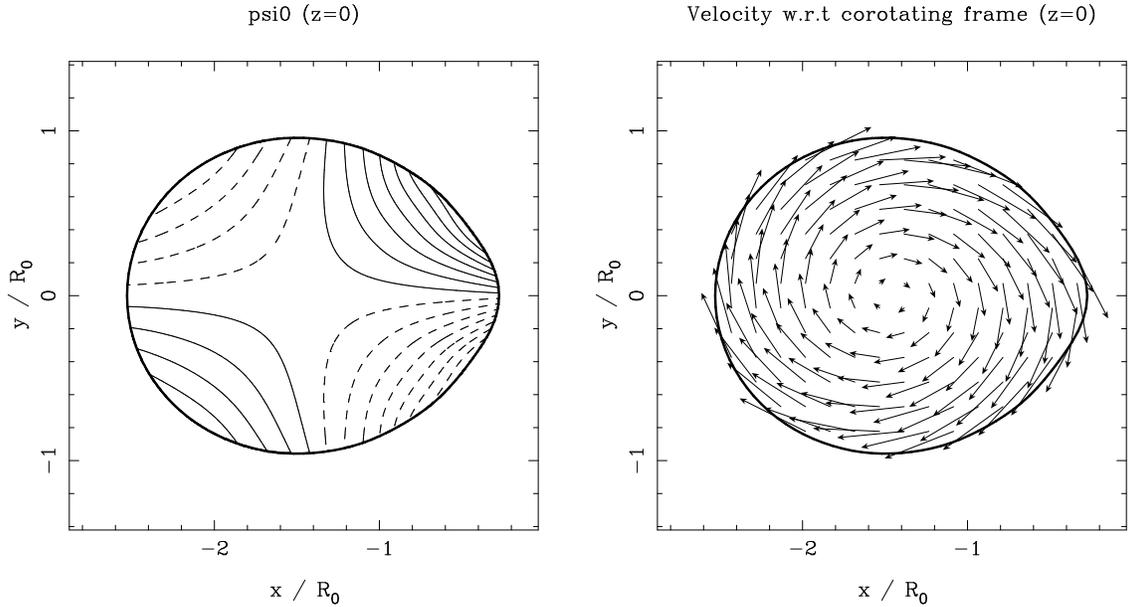

  \centerline{ \epsfig{figure=psi0_sf_ig2.eps,height=8cm}
	\hspace{0.5cm} \epsfig{figure=vs_corot_sf_ig2.eps,height=8cm} }
\vspace{0.3cm}
\caption[]{\label{fig:velo_ig2}
Same as Fig. \ref{fig:velo_ig3} but for $\gamma=2$ with $d/R_0=2.917$.}
\end{figure}

\begin{figure}
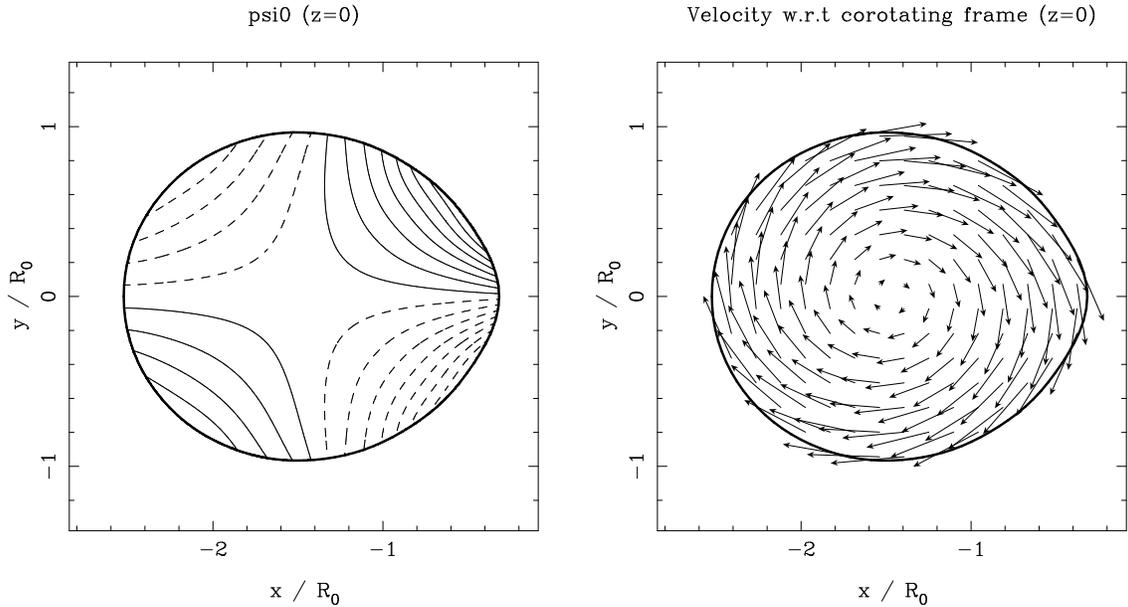

  \centerline{ \epsfig{figure=psi0_sf_ig18.eps,height=8cm}
	\hspace{0.5cm} \epsfig{figure=vs_corot_sf_ig18.eps,height=8cm} }
\vspace{0.3cm}
\caption[]{\label{fig:velo_ig18}
Same as Fig. \ref{fig:velo_ig3} but for $\gamma=1.8$ with $d/R_0=2.932$
}
\end{figure}

Figures~\ref{fig:velo_ig3} -- \ref{fig:velo_ig18} show some 
isocontours of the velocity potential, as well as the
velocity field in the co-orbiting
frame, for the polytropic indices $\gamma=3$, $2$ and $1.8$,
respectively.
In these figures, we show only one of the two star.
The companion star is located at the position symmetric with respect
to the $y-z$ plane.
The velocity potential $\Psi_0$ is defined as
\be
  \Psi_0 := \Psi - \w{W}_0 \cdot \w{r}, \label{e:psi0_def}
\ee
where 
$\w{W}_0$ is the constant translational velocity field defined as
the central value of $\w{W} := \wg{\Omega} \times \w{r}$.
Note that the vector field is tangent to the stellar surface, as it should be.

\subsection{End points of sequences: contact vs. cusp} \label{s:end}

\begin{figure}
\vspace{0.3cm}
  \centerline{ \epsfig{figure=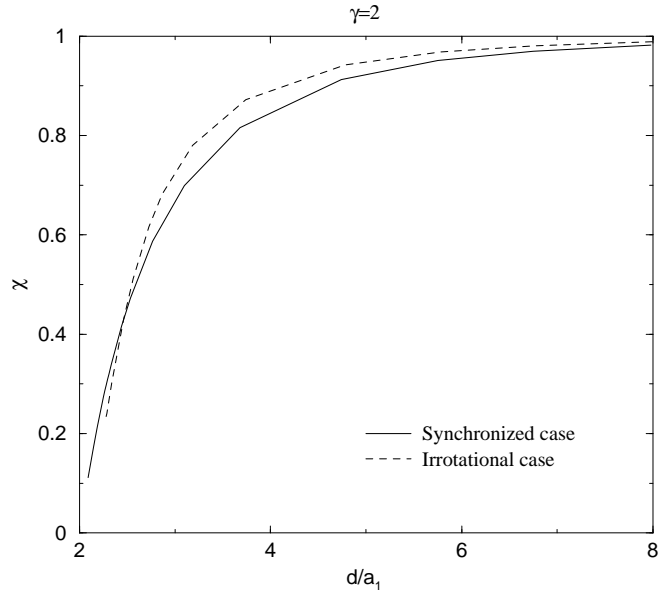,height=8cm} }
\vspace{0.3cm}
\caption[]{\label{fig:chi}
Equatorial to polar ratio of the radial derivative of enthalpy 
$\chi$ as a function of the separation $d$ 
(normalized by the radius $a_1$).
The solid and dashed lines denote the cases of synchronized
and irrotational fluid states, respectively.
}
\end{figure}

\begin{figure}
\vspace{0.3cm}
  \centerline{ \epsfig{figure=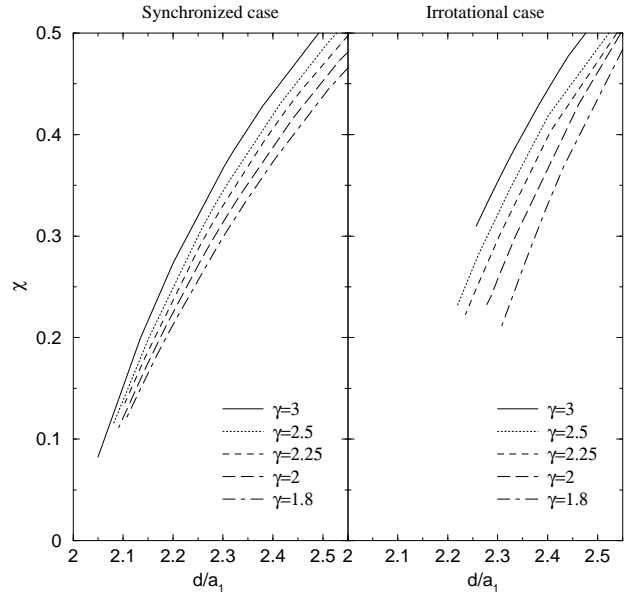,height=8cm} }
\vspace{0.3cm}
\caption[]{\label{fig:chi_magni}
Equatorial to polar ratio of the radial derivative of enthalpy $\chi$
as a function of the separation $d$ (normalized by the radius $a_1$).
The left (resp. right) panel is for the synchronized (resp. irrotational)
binaries. 
Solid, dotted, dashed, long-dashed, and dot-dashed lines denote
the cases of $\gamma=3$, $2.5$, $2.25$, $2$, and $1.8$, respectively
}
\end{figure}

An equilibrium sequence terminates by the contact between the two stars
$(d/a_1=2)$ or by a cusp at the onset of mass shedding $(\chi=0)$. 
In order to investigate which final fate occurs,
it is helpful to display a sequence in the $d/a_1-\chi$ plane.
This is done in Fig. \ref{fig:chi} where we compare the synchronized 
sequence with the irrotational one for $\gamma=2$.
It is found from this figure that the value of $\chi$
in the irrotational case
is larger than that in the synchronized case for large separations.
However, when the separation decreases below $d/a_1\sim3$,
$\chi$ in the irrotational case decreases rapidly and
becomes smaller than that in the synchronized case for $d/a_1<2.5$.
We magnify the region near the end of the sequences
in Fig.~\ref{fig:chi_magni}.
If we extrapolate the results up to the zero value of $\chi$ in the figure,
we can speculate about the final fates of sequences.
In the synchronized case, it seems that all the lines will
reach $\chi=0$ at $d/a_1=2$.
This means that the cusp does not appear before the two stars contact
with each other. 
On the other hand, in the irrotational case,
it seems that the lines may reach $\chi=0$ before $d/a_1=2$.
The values of $d/a_1$ may be
$d/a_1\sim2.1$ for $\gamma=3$ and $d/a_1\sim2.25$ for $\gamma=1.8$.
This means that the cusp may be created before contact of stars
in the irrotational case.
This behavior in the irrotational case agrees with the results of
Uryu\cite{Uryu00}.

It is worth to note here that the point where the cusp may be created
is expressed by the orbital separation divided by {\it the radius to the
companion star} ($a_1$). Then, it looks that the cusp appears later
along equilibrium sequences for higher adiabatic index,
i.e., stiffer equation of state.
However, if we express the cusp point by the orbital separation divided by
{\it the radius of a spherical star with the same mass} ($R_0$),
the order of cusp appearance will be reversed.
Although these behaviors are completely different from each other,
the origin is the same.
The fluid with lower adiabatic index is less affected by the tidal force
and slightly deformed. It results in the larger $d/a_1$ with fixing the
orbital separation $d$.

A simple explanation of the difference in $\chi$ between in the
synchronized case and in the irrotational one is as follows.
Using Eq. (\ref{e:psi0_def}) and the decomposition of the orbital motion
\be
  \wg{\Omega} \times \w{r} = \w{W}_0 + \w{W}_s,
\ee
we can rewrite Eq. (\ref{e:integ_euler_ir}) as
\be
  H +\nu -{1 \over 2} (\wg{\Omega} \times \w{r})^2
  +{1 \over 2} (\vec{\nabla} \Psi_0 -\w{W}_s)^2 ={\rm const},
  \label{e:integ_euler_ir_rw}
\ee
where $\w{W}_0$ has been defined in Sec.\ref{s:equil_seq} 
[see Eq.~(\ref{e:psi0_def})] and
$\w{W}_s$ is the spin part of the orbital motion defined sor star No.1 as
\be
  \w{W}_s := \Omega (-y_1, ~x_1,~0),
\ee
where $(x_1,y_1,z_1)$ are the Cartesian coordinate centered on star 1
(see Fig. 1 in Paper I).
In the following explanation, we pay particular attention to star 1.
Note here that the last term on the left-hand side of
Eq. (\ref{e:integ_euler_ir_rw}) is the kinetic energy
of the velocity field in the corotating frame
(see Eq. (\ref{e:velo_iner})).
Comparing Eq. (\ref{e:integ_euler_co}) with Eq. (\ref{e:integ_euler_ir_rw}),
we see that the enthalpy fields for synchronized and irrotational binaries
differ precisely by that kinetic energy.

\begin{figure}
\vspace{0.3cm}
  \centerline{ \epsfig{figure=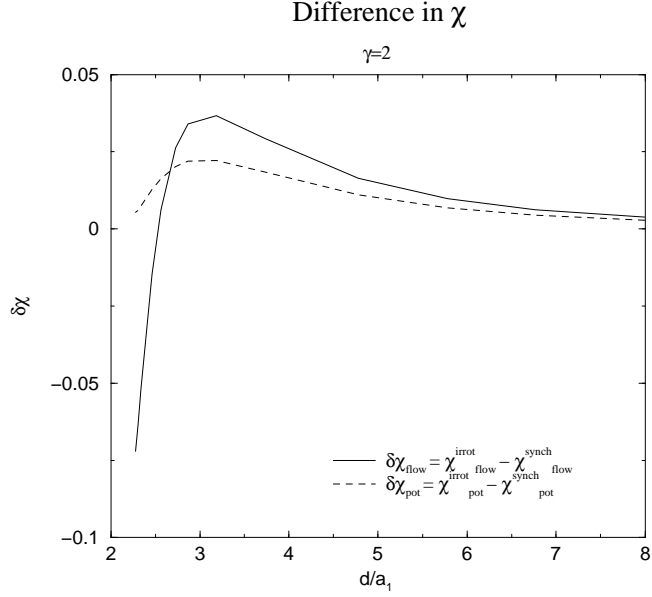,height=8cm} }
\vspace{0.3cm}
\caption[]{\label{fig:chi_diff}
Differences in $\chi$ terms between synchronized and irrotational cases
along equilibrium sequences with $\gamma=2$.}
\end{figure}

Then $\chi$ for each case becomes
\beqa
  &&\chi^{\rm synch} ={\displaystyle -\Bigl(
	{\partial \nu \over \partial x_1} \Bigr)_{\rm eq,comp}
	+\Omega^2 \Bigl( -{d \over 2} +a_1 \Bigr) \over
	\displaystyle -\Bigl( {\partial \nu \over \partial z_1}
	\Bigr)_{\rm pole}}, \\
  &&\chi^{\rm irrot} ={\displaystyle -\Bigl(
	{\partial \nu \over \partial x_1} \Bigr)_{\rm eq,comp}
	+\Omega^2 \Bigl( -{d \over 2} +a_1 \Bigr)
	-{1 \over 2} {\partial \over \partial x_1}
	(\vec{\nabla} \Psi_0 -\w{W}_s)^2 \Bigl|_{\rm eq,comp}
	\over \displaystyle -\Bigl( {\partial \nu \over \partial z_1}
	\Bigr)_{\rm pole}}. \label{e:chi_irrot}
\eeqa
Of course, due to the difference of deformation of the stars,
the gravitational potential $\nu$ for irrotational binaries
is different from that for synchronized ones,
even if we set two stars at the same orbital separation
in the binary systems.
However, the largest difference between $\chi^{\rm synch}$ and
$\chi^{\rm irrot}$ is the last term in the numerator of
Eq. (\ref{e:chi_irrot}).
We show this behavior in the following.
First, we divide $\chi$ into two parts;
\beqa
  &&\chi^{\rm synch} =\chi^{\rm synch}_{\rm pot}
	+\chi^{\rm synch}_{\rm flow}, \\
  &&\chi^{\rm irrot} =\chi^{\rm irrot}_{\rm pot}
	+\chi^{\rm irrot}_{\rm flow},
\eeqa
where
\beqa
  &&\chi^{\rm synch}_{\rm pot} ={\displaystyle -\Bigl(
	{\partial \nu \over \partial x_1} \Bigr)_{\rm eq,comp}
	+\Omega^2 \Bigl( -{d \over 2} +a_1 \Bigr) \over
	\displaystyle -\Bigl( {\partial \nu \over \partial z_1}
	\Bigr)_{\rm pole}} ~~~~~({\rm for~synchronized~binaries}), \\
  &&\chi^{\rm synch}_{\rm flow} =0, \\
  &&\chi^{\rm irrot}_{\rm pot} ={\displaystyle -\Bigl(
	{\partial \nu \over \partial x_1} \Bigr)_{\rm eq,comp}
	+\Omega^2 \Bigl( -{d \over 2} +a_1 \Bigr)
	\over \displaystyle -\Bigl( {\partial \nu \over \partial z_1}
	\Bigr)_{\rm pole}} ~~~~~({\rm for~irrotational~binaries}), \\
  &&\chi^{\rm irrot}_{\rm flow} ={\displaystyle
	-{1 \over 2} {\partial \over \partial x_1}
	(\vec{\nabla} \Psi_0 -\w{W}_s)^2 \Bigl|_{\rm eq,comp}
	\over \displaystyle -\Bigl( {\partial \nu \over \partial z_1}
	\Bigr)_{\rm pole}} ~~~~~({\rm for~irrotational~binaries}).
\eeqa
Then, we plot the differences
$\delta \chi_{\rm pot} =\chi^{\rm irrot}_{\rm pot}
-\chi^{\rm synch}_{\rm pot}$ and
$\delta \chi_{\rm flow} =\chi^{\rm irrot}_{\rm flow}
-\chi^{\rm synch}_{\rm flow}$ in Fig. \ref{fig:chi_diff}
in the $\gamma=2$ case.
Here, $\delta \chi_{\rm pot}$ denotes the difference in $\chi$
between synchronized and irrotational cases
which includes the gravitational potential force plus centrifugal force,
and $\delta \chi_{\rm flow}$ denotes one
which has relation to the internal flow in the corotating frame.

We can confirm from this figure that, at small separation,
 the difference in the gravitational
force plus centrifugal force is smaller than
that in the force related internal flow.
Therefore, we will consider only the last term in the numerator of
Eq. (\ref{e:chi_irrot}) below.

Now, we assume the following form for $\vec{\nabla} \Psi_0$:
\be
  \vec{\nabla} \Psi_0 =\Omega \Bigl( f(d,x_1,y_1,z_1),~g(d,x_1,y_1,z_1),
	~h(d,x_1,y_1,z_1) \Bigr),
\ee
where $f$, $g$ and $h$ are some scalar functions of
the orbital separation $d$ and the coordinates $(x_1,y_1,z_1)$.
Then the last term in the numerator of Eq. (\ref{e:chi_irrot})
can be calculated as
\beqa
  F :=&& -{1 \over 2} {\partial \over \partial x_1}
	(\vec{\nabla} \Psi_0 -\w{W}_s)^2, \\
  =&& -\Omega^2 \Bigl[ (f +y_1) {\partial f \over \partial x_1}
	+(g -x_1) \Bigl( {\partial g \over \partial x_1} -1 \Bigr)
	+h {\partial h \over \partial x_1} \Bigr].
\eeqa
Accordingly, the surface value at $(r_1,\theta_1,\varphi_1)=(a_1,\pi/2,0)
\Leftrightarrow (x_1,y_1,z_1)=(a_1,0,0)$ becomes
\be
  F_{\rm eq,comp} =-\Omega^2 a_1 \Bigl[ {f(a_1) \over a_1}
	{\partial f \over \partial x_1} (a_1)
	+\Bigl( {g(a_1) \over a_1} -1 \Bigr)
	\Bigl( {\partial g \over \partial x_1} (a_1) -1 \Bigr) \Bigr],
	\label{e:force_irrot}
\ee
where we have used $h(d,x_1,y_1,0)=0$ because the velocity field
is antisymmetric with respect to the $x-y$ plane.
Note that we simplify the argument list of $f$ and $g$
from $(a_1,0,0)$ to $(a_1)$. 
Let us consider the dominant terms in $f$ and $g$ which are
\beqa
  f &\simeq& \Lambda (d) y_1, \label{e:f_lam_y} \\
  g &\simeq& \Lambda (d) x_1. \label{e:g_lam_x}
\eeqa
Here $\Lambda$ is a function of the separation $d$.
The forms (\ref{e:f_lam_y}) and (\ref{e:g_lam_x}) can be justified 
from the studies by Lai, Rasio \& Shapiro\cite{LaiRS94} or 
Taniguchi \& Nakamura\cite{TanigN00b}.
Indeed, for ellipsoidal models, $\Lambda$ becomes
\be 
  \Lambda = {a_1^2 -a_2^2 \over a_1^2 +a_2^2}, \label{e:velo_ellip}
\ee
and its dependence of $d$ is $O[(d/R_0)^{-3}]$
because the dominant effect in the deviation from a spherical star
is produced by the tidal force.
From Eqs.~(\ref{e:f_lam_y})-(\ref{e:g_lam_x}) it appears that 
the first term in the brackets on the right-hand side
of Eq. (\ref{e:force_irrot}) is negligible as compared with
the second term, so that one is left with
\be
  F_{\rm eq,comp} \simeq -\Omega^2 a_1 \Bigl( {g(a_1) \over a_1} -1 \Bigr)
	\Bigl( {\partial g \over \partial x_1} (a_1) -1 \Bigr).
\ee

Taking into account Eqs.~(\ref{e:g_lam_x}) and (\ref{e:velo_ellip}), 
we can have only two types of behaviors for the function $g$:
\begin{itemize}
  \item[(i)] $\displaystyle {g(a_1) \over a_1} < 1$ and
	$\displaystyle {\partial g \over \partial x_1} (a_1) \le 1$,

  \item[(ii)] $\displaystyle {g(a_1) \over a_1} < 1$ and
	$\displaystyle {\partial g \over \partial x_1} (a_1) > 1$.
\end{itemize}
Since $g$ tends to zero for very large orbital separations,
the case (i) will occur in the earlier stage of the sequence.
In this case, $F_{\rm eq,comp}$ becomes negative value so that
it makes $\chi^{\rm irrot}$ larger value than $\chi^{\rm synch}$
\footnote{It is worth to note that since the terms
$-(\partial \nu /\partial x_1)$,
$-(\partial \nu /\partial z_1)$, and $\Omega^2 ( -d/2 +x_1)$ have
negative values, $\chi^{\rm irrot}$ becomes larger for
the negative value of $F$.}.
However, if the case (ii) actually occur, $F_{\rm eq,comp}$ can take
positive value, and $\chi^{\rm irrot}$ can be smaller than that
in the synchronized case.

\begin{figure}
\vspace{0.3cm}
  \centerline{ \epsfig{figure=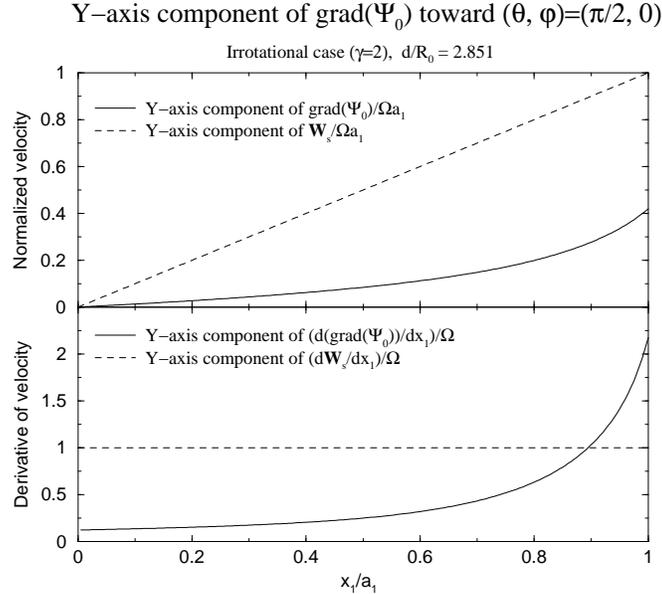,height=8cm} }
\vspace{0.3cm}
\caption[]{\label{fig:psi0}
Y-axis component of $\vec{\nabla} \Psi_0$ (i.e., the function $g$)
and $\w{W}_s$ (upper panel), and their derivatives (lower panel)
as a function of the coordinate $x_1$
for a close irrotational binary.
Solid and dashed lines are the terms concerned with $\vec{\nabla} \Psi_0$
and $\w{W}_s$, respectively.
}
\end{figure}

In Fig. \ref{fig:psi0}, the y-axis component of $\vec{\nabla} \Psi_0$,
i.e., the function $g$, and its derivative are shown as a function of
the coordinate $x_1$ (normalized by the surface value $a_1$).
This figure corresponds to the last line of the $\gamma=2$ case
in Table \ref{table2}.
We can see from this figure that the y-axis component of
$\vec{\nabla} \Psi_0$ remains smaller than that of $\w{W}_s$
throughout the interior of the star,
but the derivative near the stellar surface becomes larger than unity,
i.e., the derivative of the y-axis component of $\w{W}_s$.
This means that the case (ii) really occurs for a very close configuration.
It is worth to note here that the decrease of $\chi^{\rm irrot}$
by the term $F$ occurs only near the stellar surface.
On the contrary, the term $F$ increases $\chi^{\rm irrot}$
around the center of the star.

\begin{figure}
\vspace{0.3cm}
  \centerline{ \epsfig{figure=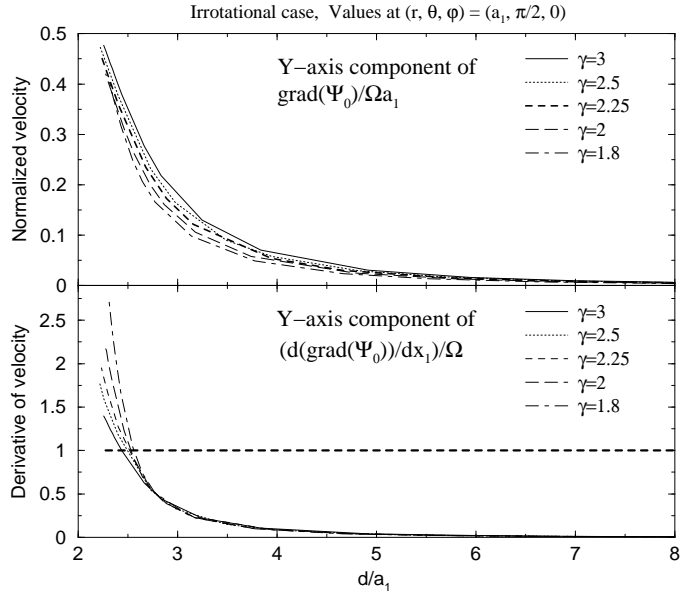,height=8cm} }
\vspace{0.3cm}
\caption[]{\label{fig:psi0_seq}
Surface value of y-axis component of $\vec{\nabla} \Psi_0$ (upper panel)
and its derivative (lower panel) along an evolutionary sequence.
Solid, dotted, dashed, long-dashed, and dot-dashed lines denote the cases
of $\gamma=3$, $2.5$, $2.25$, $2$, and $1.8$, respectively.
The thick dashed line in the lower panel is the y-axis component of
the derivative of $\w{W}_s$.
}
\end{figure}

Next, we show the surface value of y-axis component of
$\vec{\nabla} \Psi_0$ and its
derivative along an evolutionary sequence in Fig. \ref{fig:psi0_seq}.
It is found that the y-axis component of $\vec{\nabla} \Psi_0$
is smaller than that of $\w{W}_s$ throughout the sequence
even if we extrapolate the lines to $d/a_1 \sim 2.1$ for $\gamma=3$
and to $d/a_1 \sim 2.25$ for $\gamma=1.8$.
Moreover, the y-axis component of the derivative of $\vec{\nabla} \Psi_0$
becomes larger than that of $\w{W}_s$ for every $\gamma$
when the orbital separation decreases than $d/a_1 <2.5$.
This explains why $\chi^{\rm irrot}$ becomes smaller than
$\chi^{\rm synch}$ for $d/a_1 <2.5$ as we have found from Fig. \ref{fig:chi}.
We can also see from Fig. \ref{fig:psi0_seq} that the orbital separation
where the derivative of $\vec{\nabla} \Psi_0$ overcomes that of $\w{W}_s$
is larger for smaller $\gamma$.
This fact leads to the earlier appearance of a cusp for smaller $\gamma$.

\begin{figure}
\vspace{0.3cm}
  \centerline{ \epsfig{figure=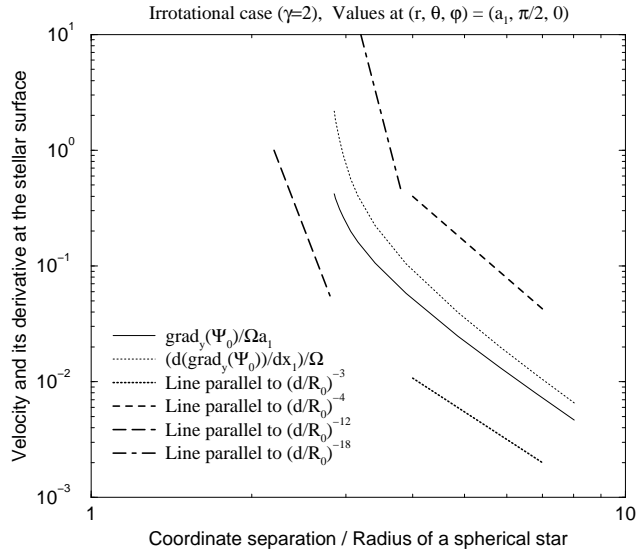,height=8cm} }
\vspace{0.3cm}
\caption[]{\label{fig:psi0_log}
Y-axis component of $\vec{\nabla} \Psi_0$ and its derivative along
an evolutionary sequence. Thick dotted, thick dashed, thick long-dashed,
and thick dot-dashed lines are reference ones.
}
\end{figure}

Finally, let us discuss the dependence on the orbital separation in
the y-axis component of $\vec{\nabla} \Psi_0$ (and also its derivative).
We present the y-axis component of $\vec{\nabla} \Psi_0$ and
its derivative along an evolutionary sequence in log-log plot
in Fig. \ref{fig:psi0_log}.
It is found that both y-axis component of $\vec{\nabla} \Psi_0$ and
its derivative behave as $O[(d/R_0)^{-3}]$
for the case of larger separation\cite{TanigN00b}.
For very close case as $d/R_0 <3$,
the y-axis component of $\vec{\nabla} \Psi_0$ becomes as $O[(d/R_0)^{-12}]$
and its derivative reaches $\sim O[(d/R_0)^{-18}]$.

\subsection{Turning points of total energy} \label{s:turning_points}

\begin{figure}
\vspace{0.3cm}
  \centerline{ \epsfig{figure=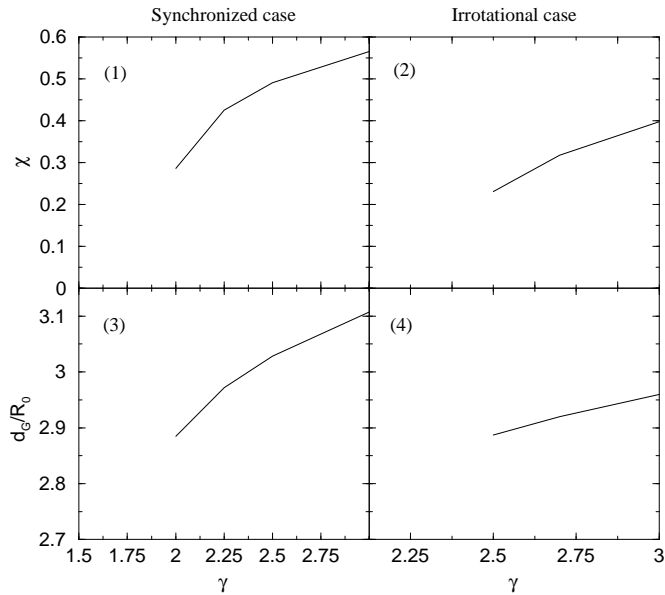,height=8cm} }
\vspace{0.3cm}
\caption[]{\label{fig:minimum}
Turning points of total energy as a function of adiabatic index $\gamma$.
Panels (1) and (3) are for synchronized binaries and
(2) and (4) are for irrotational ones.}
\end{figure}

We show values of $\chi$ and separations $d_G/R_0$
at which the total energy (and/or total angular momentum)
takes its minimum along a sequence 
as a function of the adiabatic index $\gamma$ in Fig.~\ref{fig:minimum}.
We are interested in the turning point of total energy
(and/or total angular momentum)
because it corresponds to the onset of secular instability
in the synchronized case\cite{BaumgCSST98a}
and that of dynamical instability in the irrotational one
at least for the ellipsoidal approximation\cite{LaiRS93}.
Note that the turning points in the total energy and
total angular momentum coincide.
As discussed in Sec.~\ref{s:end}, $\chi=0$ denotes the end point of
equilibrium sequences for both synchronized and irrotational cases.
If we extrapolate the results up to the zero value of $\chi$,
we can obtain the critical value of $\gamma$ below which
the turning points of the total energy does no longer exist, and
the value of the corresponding separation.

We can see from Fig. \ref{fig:minimum} that $\chi$ seems to be zero
at $\gamma =1.7 \sim 1.8$ with $d_G/R_0=2.7 \sim 2.75$
in the synchronized case and
at $\gamma =2.2 \sim 2.3$ with $d_G/R_0=2.8 \sim 2.85$
in the irrotational one.

Taking into account the appearance the cusp and
the existence of the turning point, 
we can expect that the subsequent merger process stars from
\begin{itemize}
  \item the turning point for $\gamma \ge 1.8$ (plunge ?),
  \item the contact point for $\gamma \le 1.8$
\end{itemize}
in the synchronized case, and from
\begin{itemize}
  \item the turning point for $\gamma \ge 2.3$ (plunge ?),
  \item the cusp point with mass shedding $\gamma \le 2.3$
\end{itemize}
in the irrotational one.

\section{Summary} \label{s:summary}

We have studied equilibrium sequences of both synchronized and
irrotational binary systems in Newtonian gravity
with adiabatic indices $\gamma=3,~2.5,~2.25,~2$ and 1.8.
Through the present article, we have understood
the qualitative differences of physical quantities between
synchronized binary systems and irrotational ones.
The summary of the results is as follows:
\begin{itemize}
  \item The two stars contact with each other in the synchronized case
	as an end point of equilibrium sequence.

  \item Irrotational sequences may terminate instead by a cusp point
	(mass-shedding) corresponding to a detached configuration. 

  \item The turning points of total energy appear
	in the cases of $\gamma \ge 1.8$ in the synchronized case.

  \item The turning points of total energy appear
	in the cases of $\gamma \ge 2.3$ in the irrotational case.

  \item The turning points of total energy and total angular momentum
	coincide with each other.
\end{itemize}

\acknowledgments

We thank Koji~Uryu for providing us with unpublished results from his 
numerical code. 
The code development and the numerical computations have been performed
on SGI workstations purchased thanks to a special grant from the C.N.R.S.
We warmly thank the referee for his detailed comments which helped us
to improve the quality of the article. 


\newpage

\begin{table}
\caption{Orbital angular velocity $\Omega$, total angular momentum $J$,
total energy $E$, axial ratios, relative change in central density,
equatorial to polar ratio of the radial derivative of enthalpy $\chi$, 
and virial error along equilibrium sequences in the synchronized case.
$\dagger$ denotes the minimum points of total energy
(and total angular momentum).}
 \begin{center}
  \begin{tabular}{rccccccccccl}
  \multicolumn{12}{c}{Synchronized case} \\
  $d/R_0$&$d_G/R_0$&$d/a_1$&$\bar{\Omega}$&$\bar{J}$&$\bar{E}$&
  $a_2/a_1$&$a_3/a_1$&$a_{1,{\rm opp}}/a_1$&
  $(\rho_c-\rho_{c0})/\rho_{c0}$&$\chi$&Virial error \\ \hline
  \multicolumn{12}{c}{$\gamma=3$~~($n=0.5$)} \\
           8.016&8.016&7.970&7.196(-2)&2.043&-1.172&0.9940&0.9904&0.9993&
                 -1.065(-3)&0.9838&5.145(-6) \\
           7.014&7.014&6.953&8.793(-2)&1.923&-1.180&0.9909&0.9855&0.9988&
                 -1.602(-3)&0.9754&5.054(-6) \\
           6.012&6.012&5.927&1.108(-1)&1.798&-1.191&0.9852&0.9767&0.9977&
                 -2.582(-3)&0.9602&4.920(-6) \\
           5.010&5.010&4.882&1.458(-1)&1.669&-1.205&0.9733&0.9589&0.9949&
                 -4.598(-3)&0.9288&4.707(-6) \\
           4.008&4.007&3.788&2.044(-1)&1.542&-1.224&0.9423&0.9154&0.9855&
                 -9.690(-3)&0.8496&4.326(-6) \\
           3.507&3.504&3.186&2.512(-1)&1.488&-1.235&0.9032&0.8649&0.9701&
                 -1.586(-2)&0.7526&3.994(-6) \\
  $\dagger$3.116&3.107&2.614&3.043(-1)&1.465&-1.240&0.8282&0.7781&0.9306&
                 -2.672(-2)&0.5652&3.679(-6) \\
           3.006&2.991&2.379&3.250(-1)&1.470&-1.239&0.7774&0.7245&0.8955&
                 -3.299(-2)&0.4277&4.940(-6) \\
           2.941&2.913&2.050&3.415(-1)&1.485&-1.230&0.6807&0.6299&0.8050&
                 -3.971(-2)&0.08238&2.852(-3) \\ \hline
  \multicolumn{12}{c}{$\gamma=2.5$~~($n=2/3$)} \\
           8.197&8.197&8.155&6.959(-2)&2.061&-1.137&0.9949&0.9917&0.9994&
                 -1.252(-3)&0.9841&7.434(-7) \\
           6.968&6.968&6.908&8.881(-2)&1.914&-1.147&0.9915&0.9864&0.9988&
                 -2.057(-3)&0.9738&7.260(-7) \\
           5.738&5.738&5.648&1.189(-1)&1.758&-1.161&0.9842&0.9751&0.9973&
                 -3.751(-3)&0.9518&7.012(-7) \\
           4.918&4.918&4.791&1.499(-1)&1.650&-1.173&0.9738&0.9597&0.9947&
                 -6.111(-3)&0.9211&6.747(-7) \\
           4.099&4.098&3.901&1.974(-1)&1.544&-1.189&0.9510&0.9274&0.9878&
                 -1.115(-2)&0.8553&6.245(-7) \\
           3.689&3.688&3.428&2.319(-1)&1.494&-1.198&0.9276&0.8961&0.9791&
                 -1.615(-2)&0.7895&5.889(-7) \\
  $\dagger$3.037&3.029&2.511&3.161(-1)&1.443&-1.210&0.8201&0.7701&0.9212&
                 -3.626(-2)&0.4906&8.134(-7) \\
           2.951&2.938&2.310&3.329(-1)&1.446&-1.209&0.7742&0.7224&0.8866&
                 -4.281(-2)&0.3524&3.171(-6) \\
           2.901&2.883&2.081&3.449(-1)&1.455&-1.207&0.7073&0.6568&0.8237&
                 -4.837(-2)&0.1154&2.539(-4) \\ \hline
  \multicolumn{12}{c}{$\gamma=2.25$~~($n=0.8$)} \\
           8.164&8.164&8.123&7.001(-2)&2.055&-1.108&0.9951&0.9921&0.9994&
                 -1.459(-3)&0.9835&1.180(-7) \\
           6.804&6.804&6.743&9.203(-2)&1.891&-1.119&0.9913&0.9861&0.9987&
                 -2.545(-3)&0.9709&1.143(-7) \\
           5.783&5.783&5.697&1.175(-1)&1.760&-1.131&0.9854&0.9771&0.9975&
                 -4.205(-3)&0.9516&1.118(-7) \\
           4.763&4.762&4.630&1.573(-1)&1.625&-1.147&0.9725&0.9579&0.9942&
                 -7.767(-3)&0.9098&1.055(-7) \\
           4.082&4.082&3.890&1.985(-1)&1.535&-1.161&0.9532&0.9306&0.9881&
                 -1.290(-2)&0.8494&9.882(-8) \\
           3.402&3.400&3.082&2.625(-1)&1.453&-1.177&0.9056&0.8686&0.9684&
                 -2.475(-2)&0.7058&8.189(-8) \\
           3.062&3.056&2.585&3.104(-1)&1.427&-1.183&0.8422&0.7949&0.9327&
                 -3.848(-2)&0.5178&2.051(-7) \\
  $\dagger$2.980&2.972&2.429&3.249(-1)&1.425&-1.184&0.8119&0.7623&0.9117&
                 -4.398(-2)&0.4256&8.013(-7) \\
           2.878&2.863&2.104&3.463(-1)&1.430&-1.182&0.7262&0.6762&0.8356&
                 -5.426(-2)&0.1355&4.468(-5) \\ \hline
  \multicolumn{12}{c}{$\gamma=2$~~($n=1$)} \\
           8.021&8.021&7.980&7.189(-2)&2.035&-1.061&0.9952&0.9923&0.9994&
                 -1.820(-3)&0.9819&3.154(-13) \\
           6.806&6.806&6.748&9.199(-2)&1.887&-1.072&0.9920&0.9872&0.9988&
                 -3.003(-3)&0.9698&3.234(-12) \\
           5.834&5.834&5.753&1.159(-1)&1.762&-1.083&0.9869&0.9794&0.9977&
                 -4.826(-3)&0.9511&8.484(-13) \\
           4.861&4.861&4.740&1.525(-1)&1.631&-1.098&0.9763&0.9635&0.9950&
                 -8.548(-3)&0.9126&1.530(-12) \\
           3.889&3.889&3.680&2.135(-1)&1.499&-1.119&0.9486&0.9245&0.9857&
                 -1.775(-2)&0.8160&2.063(-11) \\
           3.403&3.402&3.098&2.618(-1)&1.439&-1.131&0.9134&0.8788&0.9704&
                 -2.848(-2)&0.6990&2.283(-9) \\
           3.014&3.010&2.532&3.169(-1)&1.403&-1.140&0.8431&0.7970&0.9291&
                 -4.646(-2)&0.4725&2.259(-7) \\
  $\dagger$2.892&2.885&2.268&3.394(-1)&1.399&-1.141&0.7864&0.7373&0.8845&
                 -5.676(-2)&0.2862&4.158(-6) \\
           2.849&2.839&2.092&3.487(-1)&1.400&-1.141&0.7361&0.6878&0.8364&
                 -6.176(-2)&0.1116&1.315(-4) \\ \hline
  \multicolumn{12}{c}{$\gamma=1.8$~~($n=1.25$)} \\
           8.097&8.097&8.058&7.088(-2)&2.041&-0.9942&0.9957&0.9931&0.9994&
                 -2.087(-3)&0.9817&1.829(-9) \\
           6.701&6.701&6.643&9.416(-2)&1.869&-1.006&0.9923&0.9877&0.9988&
                 -3.712(-3)&0.9672&1.790(-9) \\
           5.584&5.584&5.498&1.238(-1)&1.722&-1.020&0.9862&0.9784&0.9974&
                 -6.502(-3)&0.9419&1.790(-9) \\
           4.746&4.746&4.622&1.580(-1)&1.606&-1.034&0.9764&0.9639&0.9947&
                 -1.081(-2)&0.9023&1.712(-9) \\
           3.909&3.908&3.710&2.118(-1)&1.489&-1.053&0.9537&0.9317&0.9867&
                 -2.029(-2)&0.8135&1.586(-9) \\
           3.350&3.349&3.043&2.677(-1)&1.415&-1.068&0.9152&0.8818&0.9692&
                 -3.462(-2)&0.6719&4.648(-9) \\
           3.071&3.069&2.651&3.062(-1)&1.384&-1.076&0.8718&0.8304&0.9439&
                 -4.821(-2)&0.5203&6.411(-8) \\
           2.932&2.928&2.405&3.296(-1)&1.372&-1.079&0.8296&0.7840&0.9135&
                 -5.863(-2)&0.3764&9.670(-7) \\
           2.828&2.822&2.104&3.496(-1)&1.367&-1.080&0.7532&0.7067&0.8440&
                 -6.939(-2)&0.1175&1.744(-5) \\
  \end{tabular}
 \end{center}
 \label{table1}
\end{table}%

\begin{table}
\caption{Orbital angular velocity $\Omega$, total angular momentum $J$,
total energy $E$, axial ratios, relative change in central density,
equatorial to polar ratio of the radial derivative of enthalpy $\chi$, 
and virial error along equilibrium sequences in the irrotational case.
$\dagger$ denotes the minimum points of total energy
(and total angular momentum).}
 \begin{center}
  \begin{tabular}{rccccccccccl}
  \multicolumn{12}{c}{Irrotational case} \\
  $d/R_0$&$d_G/R_0$&$d/a_1$&$\bar{\Omega}$&$\bar{J}$&$\bar{E}$&
  $a_2/a_1$&$a_3/a_1$&$a_{1,{\rm opp}}/a_1$&
  $(\rho_c-\rho_{c0})/\rho_{c0}$&$\chi$&Virial error \\ \hline
  \multicolumn{12}{c}{$\gamma=3$~~($n=0.5$)} \\
           8.016&8.016&7.983&7.196(-2)&2.002&-1.173&0.9940&0.9940&0.9993&
                 -7.249(-6)&0.9897&5.194(-6) \\
           7.014&7.014&6.970&8.793(-2)&1.873&-1.182&0.9909&0.9910&0.9988&
                 -1.638(-5)&0.9843&5.123(-6) \\
           6.012&6.012&5.950&1.108(-1)&1.734&-1.194&0.9851&0.9853&0.9977&
                 -4.243(-5)&0.9742&5.029(-6) \\
           5.010&5.010&4.914&1.458(-1)&1.584&-1.211&0.9729&0.9736&0.9948&
                 -1.337(-4)&0.9525&4.891(-6) \\
           4.008&4.007&3.838&2.042(-1)&1.420&-1.235&0.9408&0.9434&0.9853&
                 -5.862(-4)&0.8939&4.642(-6) \\
           3.507&3.505&3.248&2.506(-1)&1.334&-1.252&0.8994&0.9055&0.9695&
                 -1.559(-3)&0.8138&4.463(-6) \\
           3.006&2.993&2.442&3.230(-1)&1.260&-1.270&0.7622&0.7795&0.8892&
                 -6.832(-3)&0.4775&2.016(-7) \\
  $\dagger$2.976&2.960&2.346&3.296(-1)&1.259&-1.270&0.7362&0.7547&0.8686&
                 -7.981(-3)&0.3973&9.366(-6) \\
           2.956&2.936&2.257&3.346(-1)&1.260&-1.270&0.7099&0.7292&0.8458&
                 -9.095(-3)&0.3097&3.220(-5) \\ \hline
  \multicolumn{12}{c}{$\gamma=2.5$~~($n=2/3$)} \\
           8.197&8.197&8.168&6.959(-2)&2.025&-1.138&0.9948&0.9949&0.9994&
                 -7.110(-6)&0.9900&7.487(-7) \\
           6.968&6.968&6.926&8.880(-2)&1.867&-1.149&0.9914&0.9915&0.9988&
                 -1.919(-5)&0.9833&7.382(-7) \\
           5.738&5.738&5.674&1.189(-1)&1.694&-1.164&0.9841&0.9843&0.9973&
                 -6.355(-5)&0.9686&7.194(-7) \\
           4.918&4.918&4.826&1.498(-1)&1.570&-1.178&0.9736&0.9742&0.9947&
                 -1.677(-4)&0.9475&7.037(-7) \\
           4.099&4.098&3.952&1.973(-1)&1.435&-1.199&0.9502&0.9521&0.9878&
                 -5.524(-4)&0.8996&6.760(-7) \\
           3.689&3.688&3.490&2.316(-1)&1.364&-1.212&0.9260&0.9297&0.9791&
                 -1.149(-3)&0.8481&6.403(-7) \\
           3.279&3.276&2.978&2.778(-1)&1.293&-1.227&0.8777&0.8853&0.9574&
                 -2.832(-3)&0.7372&6.161(-7) \\
           2.951&2.941&2.401&3.306(-1)&1.244&-1.240&0.7680&0.7833&0.8862&
                 -7.947(-3)&0.4184&9.912(-6) \\
  $\dagger$2.902&2.887&2.219&3.415(-1)&1.240&-1.241&0.7162&0.7334&0.8415&
                 -1.023(-2)&0.2305&7.755(-5) \\ \hline
  \multicolumn{12}{c}{$\gamma=2.25$~~($n=0.8$)} \\
           8.164&8.164&8.137&7.000(-2)&2.021&-1.109&0.9951&0.9951&0.9994&
                 -7.639(-6)&0.9896&1.179(-7) \\
           6.804&6.804&6.762&9.203(-2)&1.845&-1.121&0.9913&0.9914&0.9987&
                 -2.327(-5)&0.9814&1.164(-7) \\
           5.783&5.783&5.724&1.175(-1)&1.701&-1.134&0.9854&0.9856&0.9975&
                 -6.332(-5)&0.9686&1.142(-7) \\
           4.763&4.762&4.669&1.573(-1)&1.545&-1.152&0.9723&0.9730&0.9942&
                 -2.145(-4)&0.9399&1.108(-7) \\
           4.082&4.082&3.943&1.984(-1)&1.432&-1.170&0.9528&0.9544&0.9882&
                 -5.859(-4)&0.8960&1.065(-7) \\
           3.402&3.400&3.159&2.620(-1)&1.312&-1.193&0.9043&0.9094&0.9687&
                 -2.119(-3)&0.7810&9.513(-8) \\
           3.062&3.057&2.681&3.091(-1)&1.254&-1.208&0.8401&0.8501&0.9333&
                 -5.056(-5)&0.6038&2.917(-7) \\
           2.926&2.918&2.406&3.334(-1)&1.233&-1.213&0.7806&0.7942&0.8903&
                 -8.038(-3)&0.4036&1.281(-5) \\
           2.875&2.864&2.235&3.442(-1)&1.228&-1.215&0.7333&0.7487&0.8494&
                 -1.011(-2)&0.2226&8.615(-5) \\ \hline
  \multicolumn{12}{c}{$\gamma=2$~~($n=1$)} \\
           8.021&8.021&7.995&7.189(-2)&2.003&-1.062&0.9952&0.9952&0.9994&
                 -8.767(-6)&0.9886&3.324(-12) \\
           6.806&6.806&6.768&9.198(-2)&1.845&-1.073&0.9920&0.9920&0.9988&
                 -2.382(-5)&0.9808&2.828(-12) \\
           5.834&5.834&5.780&1.159(-1)&1.708&-1.086&0.9869&0.9871&0.9977&
                 -6.136(-5)&0.9684&2.685(-13) \\
           4.861&4.861&4.779&1.525(-1)&1.560&-1.103&0.9763&0.9767&0.9950&
                 -1.912(-4)&0.9422&2.996(-13) \\
           3.889&3.889&3.743&2.134(-1)&1.397&-1.128&0.9486&0.9504&0.9859&
                 -8.118(-4)&0.8724&1.558(-11) \\
           3.403&3.402&3.182&2.614(-1)&1.311&-1.146&0.9138&0.9178&0.9712&
                 -2.058(-3)&0.7799&1.467(-9) \\
           2.965&2.962&2.561&3.240(-1)&1.232&-1.166&0.8295&0.8390&0.9536&
                 -6.198(-3)&0.5118&2.299(-6) \\
           2.917&2.912&2.462&3.328(-1)&1.224&-1.168&0.8081&0.8187&0.9042&
                 -7.255(-3)&0.4298&8.920(-6) \\
           2.851&2.844&2.278&3.457(-1)&1.214&-1.171&0.7602&0.7730&0.8628&
                 -9.293(-3)&0.2325&8.369(-5) \\ \hline
  \multicolumn{12}{c}{$\gamma=1.8$~~($n=1.25$)} \\
           8.097&8.097&8.073&7.088(-2)&2.012&-0.9951&0.9957&0.9958&0.9994&
                 -8.152(-6)&0.9885&1.846(-9) \\
           6.701&6.701&6.665&9.416(-2)&1.831&-1.008&0.9923&0.9924&0.9988&
                 -2.573(-5)&0.9791&1.823(-9) \\
           5.584&5.584&5.529&1.238(-1)&1.671&-1.023&0.9862&0.9864&0.9974&
                 -7.867(-5)&0.9624&1.868(-9) \\
           4.746&4.746&4.666&1.580(-1)&1.541&-1.039&0.9766&0.9770&0.9948&
                 -2.162(-4)&0.9355&1.805(-9) \\
           3.909&3.909&3.777&2.116(-1)&1.400&-1.061&0.9543&0.9556&0.9871&
                 -7.505(-4)&0.8722&1.747(-9) \\
           3.350&3.350&3.140&2.673(-1)&1.299&-1.082&0.9176&0.9209&0.9709&
                 -2.141(-3)&0.7629&3.631(-9) \\
           3.071&3.070&2.775&3.054(-1)&1.247&-1.095&0.8777&0.8834&0.9482&
                 -4.076(-3)&0.6312&1.613(-7) \\
           2.932&2.929&2.550&3.284(-1)&1.221&-1.102&0.8399&0.8475&0.9213&
                 -5.944(-3)&0.4843&3.355(-6) \\
           2.837&2.833&2.309&3.461(-1)&1.204&-1.107&0.7803&0.7907&0.8676&
                 -8.223(-3)&0.2118&1.018(-4) \\
  \end{tabular}
 \end{center}
 \label{table2}
\end{table}%

\end{document}